\documentclass[10pt, a4]{article}
\usepackage{geometry}
\usepackage{comment}
\usepackage[ruled]{algorithm}
\usepackage{algpseudocode}
\usepackage{amsmath, amsthm, amssymb,amsfonts}
\usepackage{bbm}
\usepackage{graphicx}
\usepackage{color}
\usepackage{booktabs}
\usepackage{caption}
\usepackage{subcaption}
\usepackage{hyperref}
\newtheorem{theorem}{Theorem}

\newtheorem{lemma}{Lemma}

\newtheorem{remark}{Remark}

\newtheorem{corollary}{Corollary}

\newcommand{\cC}{{\mathcal C}}
\newcommand{\cL}{{\mathcal L}}
\newcommand{\cT}{{\mathcal T}}
\newcommand{\cP}{{\mathcal P}}
\newcommand{\cR}{{\mathcal R}}

\newcommand{\norm}[1]{\| #1 \|}

\newcommand{\E}{\mathbb{E}}
\newcommand{\sincronia}{$\mathtt{Sincronia} \ $}

\newcommand{\blindflow}{$\mathtt{BlindFlow} \ $}
\newcommand{\philae}{$\mathtt{Philae}$}
\newcommand{\aalo}{$\mathtt{Aalo}$}
\newcommand{\bo}[1]{\mathbf{#1}}

\begin{document}

\title{Performance bounds for priority-based stochastic coflow scheduling}
\author{Olivier Brun and Balakrishna J. Prabhu, \\
LAAS-CNRS, University of Toulouse, CNRS,  31400  Toulouse, France. \\
E-mails: 	\{brun, balakrishna.prabhu\}@laas.fr.
}

\maketitle

\begin{abstract}
	We consider the coflow scheduling problem in the non-clairvoyant setting, assuming that flow sizes are realized on-line according to given probability distributions. The goal is to minimize the weighted average completion time of coflows in expectation. We first obtain inequalities for this problem that are valid for all non-anticipative order-based rate-allocation policies and define a polyhedral relaxation of the performance space of such scheduling policies. This relaxation is used to analyze the performance of a simple priority policy in which the priority order is computed by \sincronia from expected flow sizes instead of their unknown actual values. We establish a bound on the approximation ratio of this priority policy with respect to the optimal priority policy for arbitrary probability distributions of flow sizes (with finite first and second moments). Tighter upper bounds are obtained for some specific distributions.  Extensive numerical results suggest that performance of the proposed policy is much better than the upper bound.
\end{abstract}

%\begin{IEEEkeywords}
%	{Stochastic coflow scheduling, approximation ratio, Sincronia}
%\end{IEEEkeywords}

\section{Introduction}
\label{sec:introduction}

Modern data centres are increasingly used to run data-parallel computing applications such as MapReduce, Hadoop and Spark~\cite{dean2004mapreduce,zaharia2010spark}. These applications alternate between computation stages and communication stages, during which the application’s tasks exchange intermediate results using the data centre network. Each application may generate hundreds or thousands of concurrent data transfers which have to share the resources of the datacenter network not only between themselves, but also with the data flows of the other applications. Chowdhury and Stoica showed in \cite{ChowdhuryHotNet2012} that by allocating rates to all these data transfers in some appropriate way, it was possible to significantly reduce the average time spent in communication by data-parallel applications. They introduced the coflow scheduling problem, a coflow being defined as a collection of flows generated by the same parallel applicaton and its completion time being determined by the completion time of the last flow in the collection (that is, it corresponds to the time at which the next computation stage can start). The coflow scheduling problem is then to allocate rates to flows over time so that the average coflow completion time (CCT) is minimized.

Even in the clairvoyant offline setting, in which all coflows to be scheduled are initially present in the system and the precise volume of each and every flow is known, the problem is known to be NP-hard~\cite{Chowdhury2014}. This follows from the fact that it can be seen as a generalization of the concurrent open-shop scheduling problem in which resources are coupled, as transmitting a flow requires capacity on both the input and output ports. Furthermore, the coflow scheduling problem was shown to be inapproximable below a factor of $2$~\cite{Bansal2010,Sachdeva2013} in this setting. Several heuristics and approximation algorithms for clairvoyant coflow scheduling have been proposed in the last decade, see e.g. \cite{Chowdhury2014, Chen2016, Wang2017, Wang2019, chowdhury2019, Mao2018, Shi2021, Zhang2021}. At the present time, the state-of-the-art algorithm for coflow scheduling in the clairvoyant setting is \sincronia~\cite{Agarwal2018}, which was proposed by Agarwal \emph{et al.} in 2018 and provides an approximation ratio of $4$,  the best known one (see also \cite{Shafiee2018} and \cite{Ahmadi2020} for closely-related references that were published roughly concurrently with \cite{Agarwal2018}) . In addition, an attractive feature of \sincronia is that it relies on priority mechanisms which are supported by most transport layers. It is worth mentioning that \sincronia  is not restricted to the offline setting but can also efficiently handle online coflow arrivals.

Although \sincronia  provides a fairly efficient solution to the coflow scheduling problem, it requires \emph{a priori} knowledge of the flow volumes, which is difficult to obtain in practice. Therefore, some authors have  investigated the non-clairvoyant setting, where on a coflow arrival, only the number of flows and their input/output ports are revealed to the coflow scheduler, while their volumes remains unknown~\cite{Chow2015,Zhang2018,Gao2016,Liu2020,Zhang2016,Dogar2014,Bhimaraju2020}. However, none of the proposed algorithms comes with performance guarantees, with the noticeable exception of the \blindflow  algorithm which was shown in \cite{Bhimaraju2020} to be $8 \, p$-approximate, where $p$ is the maximum number of flows that any coflow can have. Given the orders of magnitude of $p$ in practice, this performance guarantee is of course much weaker than that provided by \sincronia in the clairvoyant setting.

It could be argued however that assuming absolutely no prior information on flow volumes, as was done in the above works on the non-clairvoyant setting, is excessively pessimistic. Indeed, historical data on previous executions of data-parallel applications could be exploited to get some insights on what the flow volumes could be. For instance, machine learning (ML) algorithms could be used to obtain predictions on flow volumes from historical data. The improvement of online scheduling algorithms via unreliable ML predictions is a recent line of research which has attracted a lot of studies in the last few years (see e.g.~ \cite{Purohit2018, Im2022}) and which was recently applied to non-clairvoyant coflow scheduling~\cite{Brun2023}.  Another relevant possibility, which is explored in the present paper, is to assume that the unknown flow volumes are on-line realizations of known probability distributions, which can be estimated from flow sizes data collected during previous executions.

Stochastic coflow scheduling has received little attention up to now. To the extent of our knowledge, the only work that has addressed the coflow scheduling problem in the stochastic setting is \cite{Mao2018}, in which the authors study the non-preemptive coflow scheduling problem. The objective is to minimize the weighted expected completion time of coflows. The proposed approach uses a time-indexed linear programming relaxation first introduced in \cite{Skutella2016}, and uses its solution to come up with a feasible schedule. The proposed algorithm is shown to achieve an approximation ratio of $(1+2\log L)(1+\sqrt{L} \eta_{max})(1+L \eta_{max})(1+\eta_{max}^2)$, where $L$ is the number of ports and $\eta_{max}$ is an upper bound on the coefficient of variation of processing times\footnote{The processing time of a coflow on a port is the total time required for transmitting all of its constituent flows  on that port in isolation.}. We, on the other hand, consider the preemptive case.

Our work relies on the assumption that the scheduler has knowledge of the statistics of flow volumes of the coflows. The scheduling decisions are then based purely on the mean flow volumes. A complementary line of work to ours is when the statistics are not known to the scheduler but it can obtain them by scheduling a small percentage of flows from every coflow. Once an estimation is obtained, the scheduler can apply a policy of its choice based on this estimation. This line of work was investigated in \cite{JHL2022} in which they proposed \philae, an algorithm to estimate the statistics from executing a small subset of coflows. Six different scheduling policies were compared in \philae, couple of them based on Aalo~\cite{CS2015} while some of them were new. The performance of these policies was compared numerically and no guarantees in terms of approximation ratio was provided. As mentioned in \cite{JHL2022}, "\sincronia\ is orthogonal to \philae\ $\hdots$". \philae's estimation method can be used in conjunction with \sincronia\ as the scheduling policy. Our aim is to provide an approximation ratio for \sincronia\ when it computes the priority order based on mean flow volumes whereas the work of \cite{JHL2022} proposes a method to obtain estimates of mean flow volumes in real-time but does not focus on the computation of the approximation ratio. A related sampling and estimation technique was proposed in \cite{JHLD2023} but for cluster job scheduling.

% Contributions
\subsection{Contributions}
\label{subsec:contributions}
We consider the non-clairvoyant setting in which the size of each flow is unknown to the coflow scheduler and is revealed to it only upon flow completion. Although flow sizes are not known in advance, we assume that they are drawn from known probability distributions. It is shown that for any order-based rate-allocation policy, the vector of expected completion times satisfies certain inequalities, from which a polyhedral relaxation of the performance space of such coflow scheduling policies is obtained. We use this polyhedral relaxation to analyze the performance of a priority policy in which the priority order is computed with \sincronia from expected flow sizes.  Specifically, its is shown that the approximation ratio of this policy with respect to the optimal priority policy is upper bounded by $4(1 + \sqrt{L} \sigma_{max}/\mu_{min})$ for any distribution of flow sizes with finite first and second moments, where $\mu_{min}$ is a lower bound on the expected processing times and $\sigma_{max}$ is an upper bound on their standard deviations. Tighter bounds can be obtained if we have information on the distribution of processing times. For instance, we show that for normally distributed processing times, the approximation ratio of the $\mathtt{Sincronia}$-based priority policy is upper bounded by $4(1+\sqrt{2 \log L} \eta_{max})$, where, as before, $\eta_{max}$ is an upper bound on the coefficient of variation of processing times. We also obtain asymptotic bounds for Pareto-distributed processing times.

% Organization
\subsection{Organization}
\label{subsec:organization}
The paper is organized as follows. Section \ref{sec:stochastic-coflow-scheduling} introduces the main notations used throughout the paper and presents the stochastic coflow scheduling problem that we address. In Section \ref{sec:lp-approx}, we establish a polyhedral relaxation of the performance space of order-based coflow scheduling policies and introduce our $\mathtt{Sincronia}$-based priority policy. Upper bounds on the approximation ratio of this policy with respect to the optimal priority policy are proven in Section \ref{sec:bounds-alpha}. Section \ref{sec:numerics} is devoted to numerical results. Finally, some conclusions are drawn and future research directions are discussed in Section \ref{sec:conclusion}.

% Stochastic non-preemptive coflow scheduling
\section{Stochastic coflow scheduling}
\label{sec:stochastic-coflow-scheduling}
We consider a set $\cC= \{1,2,\ldots, n \}$ of coflows, which are all present in the system at time $0$. Each coflow $k$ is a set $F_k$ of flows, where flow $j$ of coflow $k$ is characterized by its source node, its destination node and its volume $V^{k,j}$. Each coflow $k$ may also be assigned a weight $w_k$ (default weight is $1$). As for the network, we assume the \emph{big switch model}\footnote{This model abstracts out the datacenter network fabric as one big switch interconnecting servers. The underlying assumption is that the fabric core can sustain 100\% throughput and only the ingress and egress ports are potential congestion points.}~\cite{Chowdhury2014} and we let $b_\ell$ denote the capacity of port $\ell \in \cL$. We let $F_{k,\ell}$ be the set of flows of coflow $k$ that use port $\ell$ either as ingress or egress port, and denote by $C_k$ the completion time of coflow $k$, that is, the time at which its last flow finishes.

In the clairvoyant setting where flow sizes $V^{k,j}$ are known to the coflow scheduler, the goal is to find a rate allocation $r(t)$ to flows, where  $r^{k,j}(t) \in \mathbb {R}_{+}$ is the rate allocated to flow $j\in F_k$ at time $t$, which minimizes the weighted average coflow completion time $\sum_{k\in\mathcal{C}}  w_k C_k$. Mathematically, the problem can be stated as follows:

\begin{align}
\underset{r}{\mathrm{min}}\enskip & \sum_{k\in\mathcal{C}}  w_k C_k \tag{P1}\label{prob:minCCT}\\
\mathrm{s.t.}\enskip 
& \sum_{k \in \cC} \sum_{j\in F_{k,\ell}} r^{k,j}(t)\leq  b_{\ell}, \quad \forall \ell \in \cL, \forall t \in \mathcal{T}, \label{prob:minCCT-1} \\
& \int_0^{C_k} r^{k,j} (t) \mathop{dt} \geq  V^{k,j}, \quad \forall j \in F_k, \forall k \in \cC, \label{prob:minCCT-2}
\end{align}

\noindent where  $\cT$ is the time horizon. Constraint \eqref{prob:minCCT-1} expresses that, at any instant $t$, the total rate that port $\ell$ assigns to flows cannot exceed its capacity $b_\ell$. Constraint \eqref{prob:minCCT-2} ensures that the data of flows of each coflow $k$ should be completely transmitted before its completion time $C_k$. 

Here we consider the non-clairvoyant setting in which the size of each flow is unknown to the coflow scheduler and is revealed to it only upon flow completion. Although flow sizes are not known in advance, we assume that they are realized on-line according to given probability distributions. Throughout the paper, flow volumes $V^{k,j}$ are supposed to be statistically independent, with finite first and second moments. The aim is to find a rate-allocation policy that minimizes the weighted average completion time of coflows in expectation, which by linearity of expectation is $\sum_{k\in\mathcal{C}}  w_k \E \left [ C_k \right ]$. 

A rate allocation policy $r$ specifies the rate allocated to each flow at any given time in such a way that constraint \eqref{prob:minCCT-1} is satisfied. We shall specifically consider work-conserving policies which allocate a constant rate to flows in between decision epochs, where decision epochs correspond to time $0$ and to flow completion times.  If $t$ is a decision epoch, the decisions of such a policy may only depend on the "past up to time $t$", which is given by the sets of data transfers already finished or being performed at $t$, as well as on the \emph{a priori} knowledge of the input data of the problem. This includes of course the distributional information about remaining sizes of unfinished flows. In other words, admissible policies must be nonanticipative in the sense that scheduling decisions cannot depend on future information, such as the actual volumes of unfinished flows. A simple example of a work-conserving and nonanticipative scheduling policy is the round-robin (RR) policy which allocates rate $r^{k,j}(t)=\min \left \{ b_i/n_i(t), b_o/n_o(t) \right \}$ to flow $j \in F_k$ at time $t$, where $i$ and $o$ are the input and output ports of this flow, respectively, and $n_\ell(t)$ is the number of ongoing flows on port $\ell$ at time $t$. 

Among work-conserving and nonanticipative rate-allocation policies,  we restrict ourselves to \emph{order-based policies}. Such policies define a partial order on the set of flows such that one flow precedes another one if they have either the same source or the same destination. Under order-based policies, no single port is transmitting multiple flows at the same time. The analysis of order-based policies is much simpler as it is possible to leverage a well-known reduction to the Concurrent Open Shop problem~\cite{Khuller2018}. Moreover, a key advantage of these policies is that they are easily implementable by leveraging priority mechanisms supported by most transport layers. An important property of order-based policies is that  the completion order of the flows (and hence of coflows) on all ports is fixed and independent of flow sizes. In the following, we let $\cR$ be the set of order-based policies. Note that the RR policy does not belong to $\cR$ as multiple flows may be served in parallel. However, $\cR$ contains all priority policies which schedule coflows in some priority order $\pi$. Such a policy allocates a positive rate to a flow of coflow $\pi(k')$ if and only if no flows of a coflow  $\pi(k)$ such that $k<k'$ is available for transmission on the same input/output port.

The problem that we consider is therefore the following

\begin{align}
\underset{r}{\mathrm{min}}\enskip & \sum_{k\in\mathcal{C}}  w_k \E \left [ C_k(r) \right ] \tag{P2}\label{prob:minECCT}\\
\mathrm{s.t.}\enskip 
& r \in \cR, \label{prob:minECCT-1}
\end{align}

\noindent where the random variable $C_k(r)$ represents the completion time of coflow $k$ under order-based policy $r \in \cR$. In the following, we denote by $C^{OPT}$ the optimum value for this problem, that is, $C^{OPT} = \mbox{inf}_{r \in \cR} \sum_{k\in\mathcal{C}}  w_k \E \left [ C_k(r) \right ]$.

% Sincronia with expected processing times as inputs
\section{Sincronia with expected processing times as inputs}
\label{sec:lp-approx}
In this section, we establish a generic bound on the approximation ratio of \sincronia for problem \eqref{prob:minECCT} when ran with  expected processing times as inputs. Following an approach proposed in  \cite{Mohring99}, we first show in Section \ref{subsec:parallel-inequalities} that for any order-based policy $r$, the vector of expected completion times $\E \left [ C_k(r) \right ]$ satisfies certain inequalities. The latter inequalities yield a polyhedral relaxation of the performance space of order-based scheduling policies, as discussed in Section \ref{subsec:lp-relaxation}. It turns out that this polyhedral relaxation is exactly the same as the one used in \sincronia in the clairvoyant setting, except that the actual processing times are replaced by their expected values. In Section \ref{subsec:sincronia-priority-policy}, we briefly describes how \sincronia works and propose a priority policy for stochastic coflow scheduling in which rates are allocated to coflows according to the priority order computed by \sincronia from expected processing times. We obtain performance guarantees for this policy in Section \ref{subsec:greedy-rate-alloc}.

% Stochastic Parallel Inequalities
\subsection{Stochastic Parallel Inequalities}
\label{subsec:parallel-inequalities}
Let $P_{\ell,k} = \left ( \sum_{j \in F_{k,\ell}} V^{k,j} \right ) / b_\ell$ be the total transmission time of coflow $k$ at port $\ell$ in isolation. In the following, we denote by $\mu_{\ell,k}$  the expected value $\E \left [ P_{\ell,k} \right ]$ of this transmission time.

The so-called parallel inequalities are inequalities on the completion times of jobs which are valid for any feasible schedule on $m$ machines~\cite{Queyranne1993,Schulz1996SchedulingTM}. These inequalities were extended in \cite{Mohring99} to stochastic parallel-machine scheduling. We adapt the proof of Theorem 3.1 in \cite{Mohring99} to show that the same inequalities are also valid for the stochastic coflow scheduling problem.

\begin{theorem}
Let $r \in \cR$ be any order-based rate-allocation policy for coflow scheduling. Then

\begin{equation}
\sum_{k \in A} \mu_{\ell,k} \E \left [ C_k(r) \right ] \geq f_\ell(A) = \frac{1}{2} \left \{
\left ( \sum_{k \in A} \mu_{\ell,k} \right )^2 +  \sum_{k \in A} \mu_{\ell,k}^2 
\right \}, 
\label{eq:stoch-par-inequalities}
\end{equation}
\noindent for all $\ell \in \cL$ and $A \subseteq \cC$.
\label{thm:stoch-par-inequalities}
\end{theorem}
\begin{proof}
See Appendix \ref{app:proof-stoch-par-inequalities}.
\end{proof}

% LP Relaxation
\subsection{LP Relaxation}
\label{subsec:lp-relaxation}
It follows from Theorem \ref{thm:stoch-par-inequalities} that for any order-based policy $r$ the expected CCT values $\E \left [C_k(r) \right ]$ lie in the polyhedron $\cP$ defined by inequalities \eqref{eq:stoch-par-inequalities}. It implies that the following linear program provides a lower bound on $C^{OPT}$:

\begin{align}
\mbox{Minimize} & \sum_{k \in \cC} w_k C_k \label{prob:primal-lp} \tag{LP-Primal}\\
\mbox{s.t} & \nonumber \\
& \sum_{k \in A} \mu_{\ell,k} C_k \geq f_\ell(A), \quad \mbox{for all } \ell \in \cL \mbox{ and } A \subseteq \cC,  \label{prob:primal-cstr1}\\
& C_k \geq 0, \quad \mbox{for all } k \in \cC, \label{prob:primal-cstr2}
\end{align}

\noindent where $f_\ell(A)$ is defined in \eqref{eq:stoch-par-inequalities}. We thus have $C^{LP} \leq C^{OPT}$, where $C^{LP}$ is the optimum value of problem \ref{prob:primal-lp}. 

% Sincronia-based priority policy
\subsection{Sincronia-based priority policy}
\label{subsec:sincronia-priority-policy}
In order to schedule a set $\cC$ of coflows with known processing times $p_{\ell,k}$, \sincronia first computes a priority order over coflows, that is, a permutation $\pi$ of $\cC$. If $k<k'$, then the flows of coflow $\pi(k)$ are transmitted in the data centre network with strict priority over the flows of coflow $\pi(k')$. \sincronia then uses a preemptive greedy rate allocation which assigns to a single flow the full rate of its incoming and outgoing ports at any given point of time and  preserves this ordering, that is, it is guaranteed that coflows will be completed in the priority order (that is, $C_{\pi(k)} \leq C_{\pi(k')}$ if $k<k')$. 
 
 It turns out that problem \eqref{prob:primal-lp} is exactly the same as the one used in \sincronia to compute the priority order $\pi$~\cite{Agarwal2018}, except that the actual processing times $p_{\ell,k}$ have been replaced by their expected values $\mu_{\ell,k}$. A natural approach to obtain an admissible scheduling policy in the stochastic setting is therefore to compute a priority order $\pi$ using \sincronia with expected processing times $\mu_{\ell,k}$, instead of the unknown actual processing times $P_{\ell,k}$. The greedy rate allocation of \sincronia then gives an admissible rate-allocation policy for problem \eqref{prob:minECCT}. 

Before establishing performance guarantees for the above $\mathtt{Sincronia}$-based priority policy, we first briefly describe how \sincronia computes the priority order $\pi$ from expected processing times $\mu_{\ell,k}$. \sincronia uses a primal-dual algorithm in which the primal problem is \eqref{prob:primal-lp}, and the dual problem is as follows

\begin{align}
\mbox{Maximize} & \sum_{\ell \in \cL} \sum_{A \subseteq \cC} f_\ell(A) \, y_{\ell,A}  \label{prob:dual-lp} \tag{LP-Dual}\\
\mbox{s.t} & \nonumber \\
&  \sum_{A: k \in A}  \sum_{\ell \in \cL} \mu_{\ell,k} y_{\ell,A} \leq w_k, \quad \mbox{for all } k \in \cC,  \label{prob:dual-cstr1}\\
 & y_{\ell,A} \geq 0, \quad \mbox{for all } \ell \in \cL \mbox{ and } A \subseteq \cC. \label{prob:dual-cstr2}
\end{align}

The \sincronia primal-dual algorithm is shown in Algorithm \ref{alg:sincronia}. The dual variables $y_{\ell,A}$ are initialized to $0$ and the set $A$ of unscheduled coflows is initially set to $\cC$. The algorithm assigns priorities to coflows in the order of increasing priority. At each iteration $t$, the algorithm determines the bottleneck port $b$, that is, the port $\ell$ for which the total expected completion time $\sum_{k \in A} \mu_{\ell,k}$ is maximum. It then determines the coflow $k^*$ with the largest weighted expected processing time on the bottleneck. Coflow $k^*$ is assigned priority $t$ with $\pi(t)=k^*$ and it is removed from the set of unscheduled coflows $A$. Before that, the dual variable $y_{b,A}$ and the primal variable $C_{k^*}$ are set to their final values and the weights of the coflows using the bottleneck port are updated. As discussed in in \cite{Agarwal2018} , the complexity of Algorithm \ref{alg:sincronia} is $O(n^2)$.

% Sincronia algorithm
\begin{algorithm}
\caption{\sincronia primal-dual algorithm with expected processing times $\mu_{\ell,k}$ as inputs \label{alg:sincronia}}
\begin{algorithmic}[1]
\State $y_{\ell,A} \gets 0$ for all $\ell \in \cL$ and $A \subseteq \cC$.
\State $A \gets \cC$
\For{$t=n \ldots 1$}
	\State $b \gets \mbox{argmax}_{\ell \in \cL} \sum_{k \in A} \mu_{\ell,k}$ \Comment{\small{Bottleneck port}}
	\State $k^* \gets \mbox{argmin}_{k \in A} \left ( w_k/\mu_{b,k} \right )$ \Comment{\small{Largest weighted proc. time}}
	\State $C_{k^*} \gets \sum_{k \in A} \mu_{b,k}$ \Comment{\small{Set primal variable}}
	\State $y_{b,A} \gets w_{k^*}/\mu_{b,{k^*}}$ \Comment{\small{Set dual variable}}
	\State $w_k \gets w_k - w_{k^*} \mu_{b,k}/\mu_{b,{k^*}}$ for all $k \in A$ \Comment{\small{Update weights}}
	\State $\pi(t) \gets k^*$ \Comment{\small{Set priority of coflow $k^*$}}
	\State $A \gets A \setminus \{ k^* \}$ \Comment{\small{Remove $k^*$ from unscheduled coflows}}
\EndFor 
\end{algorithmic}
\end{algorithm}

The approximation ratio of Algorithm \ref{alg:sincronia} is established in \cite{Agarwal2018} and formally stated in Theorem \ref{thm:approx-ratio-sincronia} below.

\begin{theorem}
Algorithm  \ref{alg:sincronia} provides a feasible solution to problem \ref{prob:primal-lp} such that $C_{\pi(t)}^{SIN} = \max_\ell \sum_{\tau=1}^t \mu_{\ell,\pi(\tau)}$ for $t=1,2,\ldots,n$. Moreover, the cost of this solution is at most two times $C^{LP}$, that is, 

\begin{equation}
\sum_{t=1}^n w_{\pi(t)} \max_\ell \sum_{\tau=1}^t \mu_{\ell,\pi(\tau)} \leq 2 \, C^{LP} \leq 2 \, C^{OPT}.
\label{eq:sincronia-order-upper-bound}
\end{equation}

\label{thm:approx-ratio-sincronia}
\end{theorem}

% Greedy rate allocation
\subsection{Performance guarantees for the Sincronia-based priority policy}
\label{subsec:greedy-rate-alloc}
Let $\pi$ be the priority order computed by Algorithm \ref{alg:sincronia} from expected processing times. We now consider the greedy rate allocation of \sincronia (see Algorithm 2 in \cite{Agarwal2018}). We let $C_k$ be the random completion time of coflow $k$ under the greedy rate allocation when coflows are processed in the priority order $\pi$. It is proven in \cite{Agarwal2018} that $c_{\pi(t)} \leq 2 \max_\ell \sum_{\tau=1}^t p_{\ell,\pi(\tau)}$ for any fixed realization of flow volumes, so that 

\begin{equation}
\E \left [ C_{\pi(t)} \right ] \leq 2 \, \E \left [ \max_\ell \sum_{\tau=1}^t P_{\ell,\pi(\tau)} \right ] \mbox{ for all } t=1,2,\ldots,n.
\label{eq:inequality-greedy-rate-alloc}
\end{equation}

In the following, in order to combine inequalities \eqref{eq:sincronia-order-upper-bound} and \eqref{eq:inequality-greedy-rate-alloc} so as to obtain an upper bound on the approximation ratio of $\mathtt{Sincronia}$, we define the quantity $\alpha$ as follows

\begin{equation}
\alpha = \max_{t=1,\ldots,n} \frac{\E \left [ \max_\ell \sum_{\tau=1}^t P_{\ell,\pi(\tau)} \right ]}{\max_\ell \left ( \sum_{\tau=1}^t \mu_{\ell,\pi(\tau)} \right )}.
\label{eq:def-alpha}
\end{equation}

\begin{theorem}
The weighted average CCT of the $\mathtt{Sincronia}$-based priority policy is upper bounded by $4 \times \alpha$ times the optimal one.
\label{thm:main-result}
\end{theorem}
\begin{proof}
We have 
%We know that $\E \left [ C_{\pi(t)} \right ] \leq 2 \, \E \left [ \max_\ell \sum_{\tau=1}^t P_{\ell,\pi(\tau)} \right ]$ for all $t=1,\ldots,n$, so that

\begin{align}
\sum_{t=1}^n w_{\pi(t)} \E \left [ C_{\pi(t) }\right ] & \leq 2 \times \sum_{t=1}^n w_{\pi(t)} \E \left [ \max_\ell \sum_{\tau=1}^t P_{\ell,\pi(\tau)} \right ], \label{eq:bound-sin-a} \\
& \leq  2 \alpha \times \sum_{t=1}^n w_{\pi(t)} \max_\ell \left ( \sum_{\tau=1}^t \mu_{\ell,\pi(\tau)} \right ), \label{eq:bound-sin-b} \\
& \leq 4 \alpha \times C^{OPT}, \label{eq:bound-sin-c}
\end{align}

\noindent where the first inequality follows from \eqref{eq:inequality-greedy-rate-alloc}, the second one follows from \eqref{eq:def-alpha} and the last one follows from \eqref{eq:sincronia-order-upper-bound}.
\end{proof}

In the next section, we derive upper bounds on the quantity $\alpha$ in \eqref{eq:def-alpha}. In view of Theorem \ref{thm:main-result}, performance guarantees for the $\mathtt{Sincronia}$-based priority policy directly follow from these bounds.

% Bornes sur alpha
% !TEX root = ./stochastic-coflow-scheduling.tex

\section{Upper bounds on $\alpha$}
\label{sec:bounds-alpha}
We first give in Section \ref{subsec:general-bound-alpha} an upper bound on $\alpha$ that holds for any distribution of the processing times. Distribution-dependent upper bounds on $\alpha$ are derived in Section \ref{subsec:specific-bounds-alpha} for some specific distributions.

\subsection{Distribution-independent upper bound}
\label{subsec:general-bound-alpha}

For a given priority order $\theta$, let the matrix $\bo{P}^{\theta}_k = [P_{\ell,j}]$ denote the $L_k\times k$ matrix with the processing times of flows ahead of coflow $k$ on the $L_k$ ports used by coflow $k$ in the order $\theta$. For ease of notation, we use $k$ instead of $\theta(k)$. It will be apparent from the context that $\bo{P}^{\theta}_k$ denotes the $\bo{P}$ matrix for the $k$th coflow in the order $\theta$ which is in fact $\theta(k)$. Note that the matrix $\bo{P}^{\theta}_k$ can be written as $\bo{P}^{\theta}_k=\bo{M}^\theta_k + \bo{X}^\theta_k$, where $\bo{M}^\theta_k = \E\bo{P}^\theta_k$ and $\bo{X}^\theta_k$ is a zero-mean matrix. We observe from \eqref{eq:def-alpha} that 

\begin{equation}
	\alpha = \max_k \frac{\E\norm{\bo{M}^\pi_k + \bo{X}^\pi_k}}{\norm{\bo{M}^\pi_k}},
\end{equation}

\noindent where $\norm{\cdot}$ is the induced $\infty$-norm and $\pi$ is the order computed by Sincronia.  An upper bound for $\alpha$ can be obtained by taking the worst-case order and using the triangular inequality for norms,

\begin{equation}
\alpha \leq \max_\theta \max_k \frac{\E\norm{\bo{M}^\theta_k + \bo{X}^\theta_k}}{\norm{\bo{M}^\theta_k}} \leq 1 + \max_\theta \max_k \frac{\E\norm{\bo{X}^\theta_k}}{\norm{\bo{M}^\theta_k}}.
\label{eqn:alpha-ub}
\end{equation}
Thus, computing an upper-bound on $\alpha$ can be reduced to the problem of computing the expectation of the $\infty$-norm of random matrices whose elements are taken from the given distribution. These bounds will depend on the probability distribution of $X_{\ell,j}$s. For a finite $k$ but $L$ going to $\infty$, one can use extreme-value theory \cite{EKM11} which models how the maximum of a certain number of random variables behaves. When both $k$ and $L$ go to $\infty$, we refer the reader to \cite{APS23} for results on the asymptotic bounds on $\alpha$. Below, we give an upper bound on $\alpha$  that is non-asymptotic but less sharp and which only depends upon the existence of certain moments. 

Let $\mu_{min} = \min\{\mu_{\ell,j}: \mu_{\ell,j} > 0\}$ and let $p \in(1,\infty)$ be such that $m_p := \max_{\ell,j}\E|X_{\ell,j}|^p  < \infty$. That is, $\mu_{min}$ is the smallest expected flow-size among those flows that are not always $0$, and $m_p$ is the largest $p$th moment of the flow-sizes provided that it is finite.

\begin{theorem}[Upper bound]
	Let $\mu_{min} = \min_{\ell,j} \mu_{\ell,j}$ and let $ m_p := \max_{\ell,j}\E|X_{\ell,j}|^p < \infty$ for $p\in(1,\infty)$. %$\sigma_{max} = \max_{\ell,j}\sigma_{\ell,j}$. Assume $\mu_{min} > 0$ and $\sigma_{max} < \infty$. 
	Then, 
	\[
	\alpha \leq 1 + L^{1/p} \frac{m_p^{1/p}}{\mu_{min}}.
	\]
\label{theo:ub}
\end{theorem}
This result says that if $P_{\ell,j}$ have a finite $p$th moment\footnote{Note that $P_{\ell,j}$ having a finite $p$th moment is equivalent to $|X_{\ell,j}|$ having a finite $p$th moment. So, to invoke Theorem~\ref{theo:ub} it is sufficient to perform the check on $P_{\ell,j}$.}, then the upper bound on approximation ratio is of the order of $L^{1/p}$. This bound is independent of the distribution and just requires a check on whether a certain moment exists or not. This bound is useful for distribution for which it is not straightforward to obtain a bound from \eqref{eqn:alpha-ub} or by using extremal value theory.

For the proof of this theorem, we need the following simple intermediate result giving an upper bound on the second moment of the sum of absolute values of random variables in terms of an upper bound on the second moments of the individual random variables.
\begin{lemma}
	Let $Y_i$, $i=1,2,\hdots,n$ be independent random variables with $\E|Y_i|^p \leq m_p, \forall i$. Then
	\begin{align}
		\E\left(\sum_i|Y_i|\right)^p \leq n^pm_p.
	\end{align}
\label{lem:sec}
\end{lemma}
\begin{proof}
	The result is a consequence of the H\"older inequality which applied on $\sum_i|Y_i|$ gives
	\begin{align*}
		\left(\sum_i|Y_i|\right)^p &\leq n^{p-1}\left(\sum_i |Y_i|^p \right)
	\end{align*}
	Now taking expectation on both sides and using the linearity of the expectation operator and the fact that $\E|Y_i|^p \leq m_p$, we obtain the claimed result.
\end{proof}
That the above bound is tight follows by taking $Y_i = \pm 1$ with probability $0.5$.
\begin{proof}[Proof of Theorem \ref{theo:ub}]
	Let $\theta$ be an arbitrary order. We denote by $\mu^\theta_{\ell,j}$ the quantity $\mu_{\ell,\theta(j)}$. Using the standard inequality that the $\infty$-norm is a lower bound on the $p$-norm and the fact that $\mu_{min}$ is a lower bound on the means, we get  
\begin{align}
\frac{\E\norm{\bo{X}^\theta_k}}{\norm{\bo{M}^\theta_k}}
&=
 \frac{\E \max_\ell\sum_{j=1}^k |X^\theta_{\ell,j}|}{\max_\ell\sum_{j=1}^k \mu^\theta_{\ell,j}} \\
 &\leq 
 \frac{\E\left(\sum_\ell\left(\sum_{j=1}^k|X^\theta_{\ell,j}|\right)^p\right)^{1/p}}{ k\mu_{min}} \\
 &\leq
 \frac{\left(\sum_\ell\E\left(\sum_{j=1}^k|X^\theta_{\ell,j}|\right)^p\right)^{1/p}}{ k\mu_{min}}\label{eq:17}\\
  &\leq 
 \frac{\left(\sum_\ell k^p m_p\right)^{1/p}}{k\mu_{min}} = L^{1/p}\frac{m_p^{1/p}}{\mu_{min}},\label{eq:18}
%  \leq \frac{L\sigma}{\mu}.
\end{align}
where \eqref{eq:17} follows from Jensen's inequality and linearity of expectation, and \eqref{eq:18} follows from Lemma~\ref{lem:sec} and the fact that $\E|X_{\ell,j}|^p = m_p$.

The claimed results then follows from \ref{eqn:alpha-ub} since the right-hand side of \eqref{eq:18} is independent of $\theta$ and $k$.
\end{proof}
\begin{corollary}[$p=2$]
	For $p=2$, the upper bound becomes
	\begin{align}
	\alpha \leq 1 + L^{1/2} \frac{\sigma_{max}}{\mu_{min}}
	\label{eq:ubpareto} 	
	\end{align}
where $\sigma_{max} = \max_{\ell,j}\sigma_{\ell,j}$ is the maximum standard deviation of the flow sizes $P_{\ell,j}$.
\end{corollary}
This bound in Theorem~\ref{theo:ub} is consistent with \sincronia in the sense that if the flow sizes are deterministic, then $ |X|^p = 0$ for all $p > 1$, and we recover the approximation factor of \sincronia in the clairvoyant setting.

\subsection{Asymptotically matching lower bound}
\label{subsec:lower-bound-alpha}
To show that the upper bound in Theorem~\ref{theo:ub} is asymptotically tight, we shall assume that flow volumes are Pareto random variables and give asymptotic bounds as $L_k\to\infty$. 

In this section as well as in Section~\ref{subsec:specific-bounds-alpha}, to simplify the notation, we shall assume, without loss of generality that the order $\theta$ is the ascending order. The results hold for any fixed order $\theta$ since the distribution of the flow volumes are independent of the order and flow volumes will be assumed to be independent and identically distributed.

Let $k$ be finite and $P_{\ell,j}$ be independent and identically distributed random variables with shape parameter $\zeta>1$. This last condition is required for the mean of these random variables to be finite. We will need the fact that for the $p$th moment of $P_{\ell,j}$  to exist (and hence be able to use the result of Theorem~\ref{theo:ub}), $\zeta$ has to be strictly larger than $p$. 

A Pareto random variable of shape parameter $\zeta$ belongs to the class of regularly varying random variables of index $\zeta$ (see \cite{Mik99} and references therein for the definition and properties of the class of regularly varying random variables).

From the closure property of regularly varying random variables \cite{Mik99}, $\sum_{j=1}^k P_{\ell,j}$ is again a regularly varying random variable of index $\zeta$ (which is the shape parameter of the constituent Pareto random variables). Further, both $P_{\ell,j}$ and $\sum_{j=1}^k P_{\ell,j}$ belong to the maximum domain of attraction of the Fr\'echet distribution of index $\zeta$. 

For this class of distributions, the maximum grows as $L^{1/\zeta}$ \cite{Mik99}, that is,
$$
\max_{\ell} \frac{\sum_{j=1}^k P_{\ell,j}}{\sum_{j=1}^k\mu_{\ell,j}}=  \Theta(L^{1/\zeta}).
$$
For a Pareto random variable with finite $p$th moment, $\zeta$ has to be greater than $p$. To get close to the upper bound in Theorem~\ref{theo:ub}, one can take then take an $\zeta > p$ but as close to $p$ as desired. This results in a small gap to the upper bound in Theorem~\ref{theo:ub} since we cannot set $\zeta=p$. We do not know how to bridge this gap either by improving the upper bound or by exhibiting a sequence of random variables with finite $p$th moment that have a growth rate of exactly $L^{1/p}$. 
The finite $p$th moment requirement is needed since the bound of Theorem~\ref{theo:ub} is obtained under that assumption. To keep the comparison fair, we need to take $\zeta$ strictly larger than $p$ and incur a small gap in the two bounds.

% Normally distributed processing times
\subsection{Distribution-dependent tighter upper bounds}
\label{subsec:specific-bounds-alpha}
The bounds on $\alpha$ in Sections~\ref{subsec:general-bound-alpha}~and~\ref{subsec:lower-bound-alpha} are proportional to $L^{1/p}$ and only require the existence of the $p$th moment of the flow processing times. In practice, the processing times could be better behaved with existence of exponential moments. We now investigate how tighter bounds can be  obtained when exponential moments (rather than just the $p$th moment) of $P_{\ell,j}$ are finite.

% Processing times following a Gamma distribution
\subsubsection*{Gamma distributed processing times}
We start with distributions with  exponential tails, that is, $\mathbb{P}(X>t) \sim \exp(-ct)$. For these distributions, the moment generating function has a non-zero radius of convergence. We focus on gamma-distributed random variables that can also be seen as a sum of a finite number of exponentially distributed random variables.

\begin{lemma}
	Let $P_{\ell,j}$ be independent and follow a gamma distribution with scale parameter $\beta_{\ell,j}$ and shape parameter $s_{\ell,j}$. 
	Then,
	 \begin{equation}
	 	\alpha \leq c + \log\left(c + \sqrt{c^2-1}\right).
	 	\label{eq:ubgamma}
	 \end{equation} 
	 where $c=1 + \gamma \log(L)$, $\gamma = \left(\frac{\sigma_{max}}{\mu_{min}}\right)^2$ with $\sigma^2_{max}= \max_{\ell,j} s_{\ell,j}\beta^2_{\ell,j}$ is the maximum variance of the flows and 
	 $\mu_{min} =  \min_{\ell,j} s_{\ell,j}\beta_{\ell,j}$ is the smallest non-zero mean of flow-size.
	\label{lem:bound-on-bottleneck-gamma}
\end{lemma}
\begin{proof}
See Appendix~\ref{pr:lem3}.
\end{proof}
%The proof is given in appendix~\ref{pr:lem3}.

Combining  Theorem \ref{thm:main-result} and Lemma \ref{lem:bound-on-bottleneck-gamma}, we have the following result regarding the approximation ratio of \sincronia for gamma-distributed processing times.

\begin{corollary}
	If all random variables $P_{\ell,j}$ are independent and gamma-distributed, the approximation of \sincronia with expected processing times as inputs is $4(c + \log\left(c + \sqrt{c^2-1}\right))$.
	\label{cor:gamma-app-ratio}
\end{corollary}

\begin{remark}
	This bound on $\alpha$ does not directly depend upon $\eta_{max}$, which is an upper bound on the coefficient of variation of the processing times. However, if $\eta_{max} = 0$,  then $\gamma = 0$ which leads to $\alpha=1$. We thus recover the approximation ratio of $4$ of \sincronia in the clairvoyant setting. 
\end{remark}

Note in particular that for large number of ports, that is large $L$, this upper bound is roughly $\Theta(\log(L))$. The upper bound therefore increases only slowly with the number of ports in contrast to the $\Theta(L^{1/p})$ growth for the Pareto distribution in the previous subsection.

\subsubsection*{Normally distributed processing times}
We now obtain an upper bound on $\alpha$ when the processing times $P_{\ell,j}$ are normally distributed. 

\begin{lemma}
	If all random variables $P_{\ell,j}$ are independent and follow a normal distribution with mean $\mu_{\ell,j}$ and coefficient of variation $\eta_{\ell,j}$, then $\alpha \leq 1 + \sqrt{2 \log(L)} \eta_{max}$, where $L = \left | \cL \right |$ is the number of ports and $\eta_{max}= \max_{\ell,j} \eta_{\ell,j}$.
	
	\label{lem:bound-on-bottleneck}
\end{lemma}
\begin{proof}
See appendix~\ref{pr:lem:bound-on-bottleneck}
\end{proof}

As a direct consequence of Theorem \ref{thm:main-result} and Lemma \ref{lem:bound-on-bottleneck}, we have the following result regarding the approximation ratio of Sincronia for normally distributed processing times.

\begin{corollary}
	If all random variables $P_{\ell,j}$ are independent and and follow a normal distribution, the approximation ratio of the $\mathtt{Sincronia}$-based priority policy is $4 \left \{ 1 + \sqrt{2 \log(L)} \eta_{max} \right \}$.
	\label{cor:normal-app-ratio}
\end{corollary}
% \sincronia with expected processing times as inputs

Note in particular that when $\eta_{max}=0$, we retrieve the approximation ratio of \sincronia in the clairvoyant setting. Further, the upper bound only grows as $\Theta(\sqrt{\log(L)})$ which is even slower than the $\Theta(\log(L)$ growth for the gamma distribution. This is to be expected since the normal distribution has a lighter tail than the gamma distribution.

% Numerical Results
\section{Numerical Results}
\label{sec:numerics}
% !TEX root = ./stochastic-coflow-scheduling.tex

In this section, we compare the average CCTs obtained with the Sincronia-based priority policy against that of the clairvoyant Sincronia algorithm and the lower bound $C^{LP}$ on random problem instances. We also include in the comparison the round robin scheduling policy and a random order policy.

The random instances are generated using a coflow workload generator based on a real-world trace with $526$ coflows. The trace was obtained from a one-hour long run of Hive/MapReduce jobs on a $3000$-machine cluster with $150$ racks at Facebook~\cite{Chowdhury2014}. The workload generator allows upsampling the Facebook trace to desired number of ports $L$ and coflows $n$, while still keeping workload characteristics similar to the original Facebook trace. The original workload generator, which is publicly available~\cite{Sin-Rep}, was slightly modified in order to be able to generate flow sizes following a target probability distribution (e.g., a normal or Gamma distribution with specified mean and standard deviation).

In order to compare the performance of scheduling algorithms under various conditions, we generate multiple configurations by varying the following parameters:
\begin{itemize}
	\item The number of ports is always assumed to be even (the first $L/2$ ports are input ports, while the other ports are output ports) and can take values between $4$ and $40$.%the following values: $L=4, 8,16,32$.
	%\item The number of coflows $n$ is, among other values, either $n=L/2$, $n=L$ or $n=3 \, L/2$.
	\item When the flow sizes follow a specific target distribution, the mean flow size is $10$. The coefficient of variation will take different values depending on the size of the instance and will be specified in the relevant place. Otherwise, flow sizes are drawn from the empirical distribution of the Facebook trace.
	\item The weights $w_k$ of all coflows are set to $1$.  
\end{itemize}

For each possible configurations of these parameters, we randomly generate problem instances. Note that in the non-clairvoyant setting, a problem instance specifies the set of ports and their capacities as well as the set of coflows, with the respective flows of each coflow, but does not specify the actual flow sizes which are unknown to a non-clairvoyant coflow scheduler. We therefore apply the following procedure for evaluating the performance of coflow scheduling algorithms:
\begin{enumerate}
\item We first generate $100$ problem instances with the specified number of ports and number of coflows by randomly generating the set of flows $F_k$ of each coflow $k$. These problem instances do not specify the actual flow sizes and thus represents the information available to a non-clairvoyant coflow scheduler.
\item For each problem instance, we randomly generate $1000$ realizations of flow sizes according to the specified random distribution. Given a scheduling algorithm, the average CCTs obtained with these traffic realizations are averaged to estimate the expected average CCT of the algorithm over the corresponding problem instance.
\end{enumerate} 

% Small instances
\subsection{Small instances}
\label{subsec:results-small-instances}
We first consider small instances for which the lower bound $C^{LP}$ can be computed using Gurobi  as linear-programming solver. This lower bound is directly computed for each of the $100$ problem instances as this does not require the actual flow sizes, but only their mean values. For each problem instance, we also estimate the expected average CCT  $C^{ALG}$ obtained with the clairvoyant Sincronia (CL), with the non-clairvoyant Sincronia-based priority policy (NC), with round-robin (RR) and with a random order policy (RO) by averaging over the $1000$ traffic realizations for this problem instance. 

We present below the ratios $C^{ALG}/C^{LP}$ obtained for three different random distributions of flow sizes (Gamma, Normal and Pareto) as well as for flow sizes following the empirical flow size distribution in the Facebook trace. These ratios provide upper bounds on the empirical approximation ratios of the considered algorithms.

% Gamma distribution
\subsubsection*{Gamma distribution}
Let us first consider the results obtained with a Gamma distribution for flow volumes. We consider three different coefficients of variations: $\eta=0.5$, $\eta=1$ and $\eta=2$. 

The mean, standard deviation as well as the first and third quartiles of the ratios $C^{ALG}/C^{LP}$ are reported in Tables \ref{tab:LP-Gamma_C0.5}, \ref{tab:LP-Gamma_C1} and \ref{tab:LP-Gamma_C2} for flow sizes following a Gamma distribution with $\eta=0.5, 1$, and $2$, respectively. Column UB in these tables shows the theoretical upper bound on $C^{NC}/C^{LP}$ computed using \eqref{eq:ubgamma}. For $\eta=0.5$, we observe that for all considered configurations of the numbers of ports and coflows, non-clairvoyant Sincronia performs almost as well as the clairvoyant one. We also note that the variability of the ratio $C^{ALG}/C^{LP}$ is also quite low for both algorithms, indicating that they perform very well for the vast majority of our random problem instances. We also observe that the ratio $C^{RR}/C^{LP}$ has a much higher mean and standard deviation, which shows that the round-robin allocation performs poorly as compared to our NC priority policy when the variability of flow sizes is low.  When the variability of the flow sizes increases (i.e., $\eta=1$ or $\eta=2$), the performance of the non-clairvoyant Sincronia degrades as compared to the clairvoyant one. The variability of the ratio  $C^{NC}/C^{LP}$ is also much higher that that observed in the case $\eta=0.5$. Nevertheless, the performance of  NC remains fairly close to that of CL, and much better than that obtained with RR. For all three values of $\eta$, the upper bound UB is pessimistic with values much bigger than $4$ while the empirical ratios are between $1$ and $2$.

\begin{table}
\begin{center}
\begin{scriptsize}
\begin{tabular}{l c c cccc cccc cccc}
\toprule
  &  & UB & \multicolumn{4}{c}{CL} & \multicolumn{4}{c}{NC} & \multicolumn{4}{c}{RR} \\
\cmidrule(lr){3-3}\cmidrule(lr){4-7}\cmidrule(lr){8-11}\cmidrule(lr){12-15}
$L$ & $n$ &    & mean  & std & Q1  & Q3  & mean  & std & Q1  & Q3 & mean  & std & Q1  & Q3\\
\midrule
4 & 4 &  8.6 & 1.01  &  0.06  &  0.97  &  1.06   & 1.08  &  0.06  &  1.02  &  1.11  &1.55  &  0.12  &  1.48  &  1.67 \\
8 & 4 &  10  & 1.07  &  0.06  &  1.03  &  1.11    & 1.11  &  0.08  &  1.06  &  1.14 & 1.71  &  0.19  &  1.60  &  1.83 \\
8 & 8 &  10 & 1.05  &  0.05  &  1.01  &  1.08    & 1.09  &  0.05  &  1.06  &  1.13 & 2.08  &  0.17  &  2.01  &  2.19 \\
16 & 8 & 11.2 &  1.09  &  0.04  &  1.06  &  1.12  & 1.11  &  0.05  &  1.08  &  1.15  & 2.23  &  0.18  &  2.10  &  2.34 \\
16 & 16 & 11.2 &  1.09  &  0.03  &  1.06  &  1.11 & 1.12  &  0.05  &  1.08  &  1.15  & 2.80  &  0.20  &  2.66  &  2.94 \\
32 & 16 &  12.4 & 1.10  &  0.03  &  1.08  &  1.13 & 1.12  &  0.04  &  1.08  &  1.15  & 2.92  &  0.24  &  2.78  &  3.03 \\
\bottomrule
\end{tabular}
\end{scriptsize}
\caption{$C^{ALG}/C^{LP}$ for flow sizes following a Gamma distribution with $\eta=0.5$. \label{tab:LP-Gamma_C0.5}}
\end{center}
\end{table}

\begin{table}
\begin{center}
\begin{scriptsize}
\begin{tabular}{l c c cccc cccc cccc}
\toprule
  &  & UB & \multicolumn{4}{c}{CL} & \multicolumn{4}{c}{NC} & \multicolumn{4}{c}{RR} \\
\cmidrule(lr){3-3}\cmidrule(lr){4-7}\cmidrule(lr){8-11}\cmidrule(lr){12-15}
$L$ & $n$ &    & mean  & std & Q1  & Q3  & mean  & std & Q1  & Q3 & mean  & std & Q1  & Q3\\
\midrule
4 & 4 & 15.6 & 0.96  &  0.09  &  0.90  &  1.02   & 1.14  &  0.09  &  1.09  &  1.20 & 1.38  &  0.15  &  1.28  &  1.49 \\
8 & 4 &  19.5 & 1.10  &  0.10  &  1.03  &  1.18   & 1.20  &  0.11  &  1.13  &  1.28 & 1.63  &  0.21  &  1.50  &  1.78 \\
8 & 8 &  19.5 & 1.04  &  0.08  &  1.00  &  1.10   &1.18  &  0.08  &  1.14  &  1.23 & 1.90  &  0.19  &  1.77  &  2.03 \\
16 & 8 &  23.1 & 1.13  &  0.06  &  1.10  &  1.18 & 1.21  &  0.07  &  1.17  &  1.26 & 2.17  &  0.19  &  2.05  &  2.29 \\
16 & 16 & 23.1 &  1.12  &  0.04  &  1.09  & 1.15 & 1.20  &  0.06  &  1.16  &  1.24 & 2.65  &  0.18  &  2.51  &  2.76 \\
32 & 16 & 26.6 & 1.14  &  0.04  &  1.11  &  1.18 & 1.18  &  0.05  &  1.15  &  1.22 & 2.84  &  0.22  &  2.70  &  2.95 \\
\bottomrule
\end{tabular}
\end{scriptsize}
\caption{$C^{ALG}/C^{LP}$ for flow sizes following a Gamma distribution with $\eta=1$.\label{tab:LP-Gamma_C1}}
\end{center}
\end{table}

\begin{table}
\begin{center}
\begin{scriptsize}
\begin{tabular}{l c c cccc cccc cccc}
\toprule
  &  & UB & \multicolumn{4}{c}{CL} & \multicolumn{4}{c}{NC} & \multicolumn{4}{c}{RR} \\
\cmidrule(lr){3-3}\cmidrule(lr){4-7}\cmidrule(lr){8-11}\cmidrule(lr){12-15}
$L$ & $n$ &    & mean  & std & Q1  & Q3  & mean  & std & Q1  & Q3 & mean  & std & Q1  & Q3\\
\midrule
4 & 4 &  36.4 & 0.89  &  0.11  &  0.81  &  0.98  & 1.28  &  0.13  &  1.21  &  1.36  & 1.14  &  0.17  &  1.02  &  1.27  \\
8 & 4 &  49.0 & 1.18  &  0.18  &  1.03  &  1.31  & 1.43  &  0.17  &  1.33  &  1.52  & 1.54  &  0.27  &  1.35  &  1.74  \\
8 & 8 &  49.0 & 1.05  &  0.12  &  0.96  &  1.13  & 1.41  &  0.12  &  1.32  &  1.50  & 1.63  &  0.23  &  1.47  &  1.80  \\
16 & 8 & 61.1 & 1.26  &  0.10  &  1.19  &  1.33  & 1.49  &  0.10  &  1.41  &  1.55  & 2.09  &  0.23  &  1.93  &  2.25  \\
16 & 16 & 61.1 &  1.20  &  0.07  &  1.16  &  1.24  & 1.46  &  0.08  &  1.40  &  1.51  & 2.40  &  0.19  &  2.27  &  2.55  \\
32 & 16 &  73.0 & 1.27  &  0.07  &  1.21  &  1.32  & 1.40  &  0.08  &  1.35  &  1.45  & 2.76  &  0.22  &  2.60  &  2.89  \\
\bottomrule
\end{tabular}
\end{scriptsize}
\caption{$C^{ALG}/C^{LP}$ for flow sizes following a Gamma distribution with $\eta=2$.\label{tab:LP-Gamma_C2}}
\end{center}
\end{table}

\begin{figure}
     \centering
     \begin{subfigure}[b]{0.45\textwidth}
         \centering
         \includegraphics[width=\textwidth]{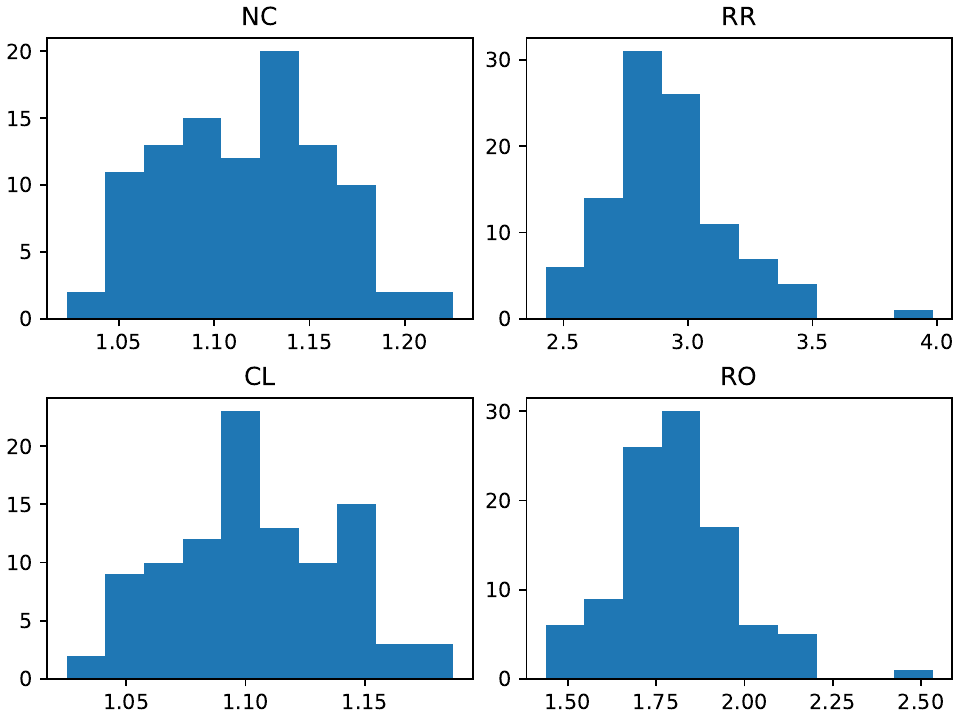}
         \caption{$\eta=0.5$}
         \label{fig:P32-N16-Gamma_C0.5}
     \end{subfigure}
     \hfill
     \begin{subfigure}[b]{0.45\textwidth}
         \centering
         \includegraphics[width=\textwidth]{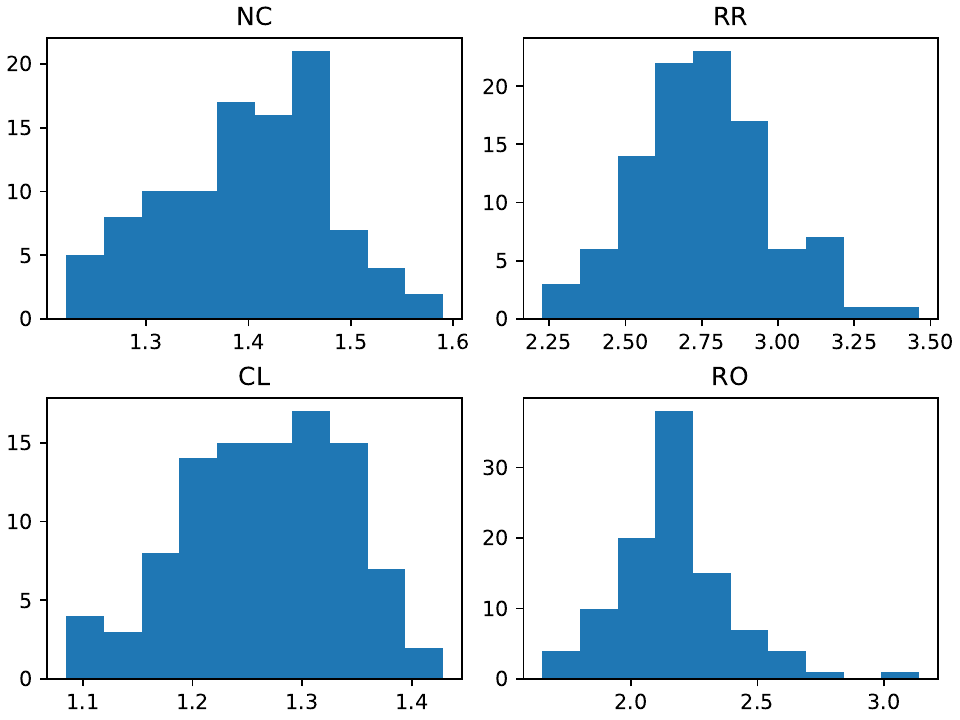}
         \caption{$\eta=2$}
         \label{fig:P32-N16-Gamma_C2}
     \end{subfigure}
     \caption{Histograms of the ratios $C^{ALG}/C^{LP}$ for CL, NC, RR and a random-order policy (RO) for $100$ random problem instances with $L=32$ ports and $n=16$ coflows. Flow sizes follow a Gamma distribution with either (a) $\eta=0.5$ or (b) $\eta=2$.\label{fig:P32-N16-Gamma}}
 \end{figure}

For a more detailed analysis, we also plot the histograms of the ratios $C^{ALG}/C^{LP}$ obtained for the $100$ random problem instances in the case $L=32$ and $n=16$ in Figure \ref{fig:P32-N16-Gamma_C0.5} and Figure \ref{fig:P32-N16-Gamma_C2} for $\eta=0.5$ and $\eta=2$, respectively. For  $\eta=0.5$, clearly the histograms obtained for $ALG=CL$ and $ALG=NC$ are very similar. The values obtained with RR are much higher, and even higher than those obtained with a random priority order.  For $\eta=2$, as compared to the histogram for $CL$, the most probable values for $NC$ are slightly shifted to the right by $+0.15$.  As in the case $\eta=0.5$, the values obtained with RR are much higher, and even higher than those obtained with a random priority order.

\begin{figure}
     \centering
	\includegraphics[width=0.6\textwidth]{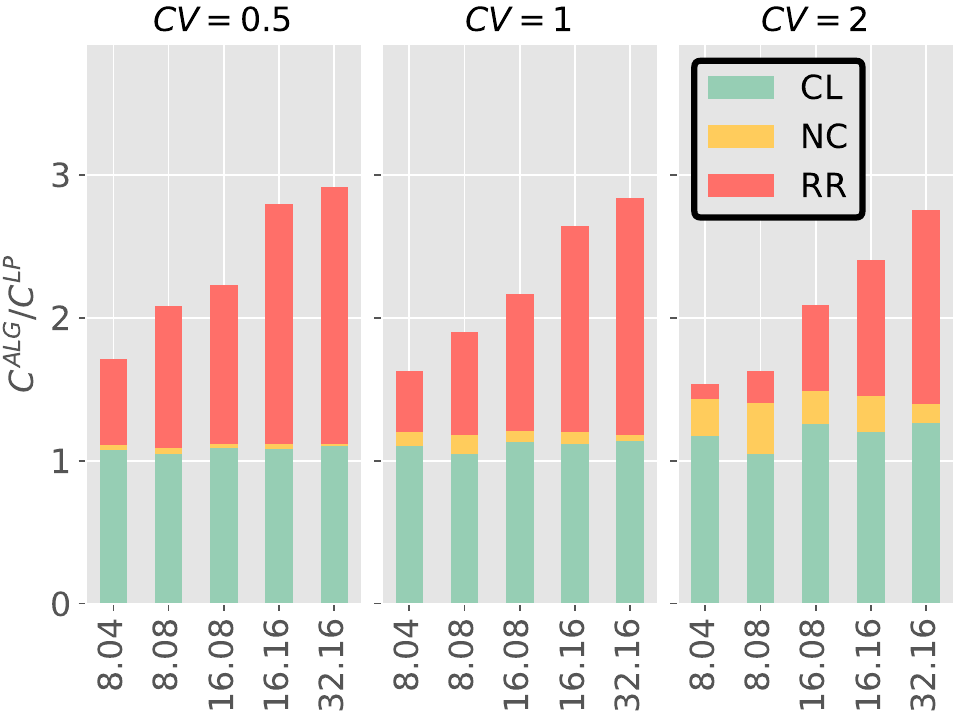}
     \caption{Average values of $C^{ALG}/C^{LP}$ for flow sizes following a Gamma distribution with $\eta=0.5$ (left), $\eta=1$ (middle) or $\eta=2$ (right). The number of ports $L$ and the number $n$ of coflows are indicated on the $x$-axis as $L.n$.\label{fig:gamma-means}}
 \end{figure}

Figure \ref{fig:gamma-means} summarizes the above results. It shows the average values (over $100$ random instances) obtained for CL, NC, and RR. We first note that CL provides near-optimal results in all cases. For $\eta=0.5$, NC performs almost as well as CL and much better than RR. For $\eta=2$, there is a noticeable performance degradation as compared to CL, but the proposed algorithm still outperforms significantly RR. We also note that the gap between CL and NC decreases as the number of ports increases.

% Normally distributed flow volumes
\subsubsection*{Normally distributed flow volumes}
We now consider the results obtained for normally-distributed flow volumes. As before, we consider three different coefficients of variations: $\eta=\frac{1}{2}$, $\eta=1$ and $\eta=2$. The mean, standard deviation as well as the first and third quartiles of the ratios $C^{ALG}/C^{LP}$ are reported in Tables \ref{tab:LP-Normal_C0.5}, \ref{tab:LP-Normal_C1} and \ref{tab:LP-Normal_C2} for $\eta=\frac{1}{2}$, $\eta=1$ and $\eta=2$, respectively. Column UB in these tables is the theoretical upper bound on $C^{NC}/C^{LP}$ computed using Cor.~\ref{cor:normal-app-ratio}. 

\begin{figure}
     \centering
     \begin{subfigure}[b]{0.45\textwidth}
         \centering
         \includegraphics[width=\textwidth]{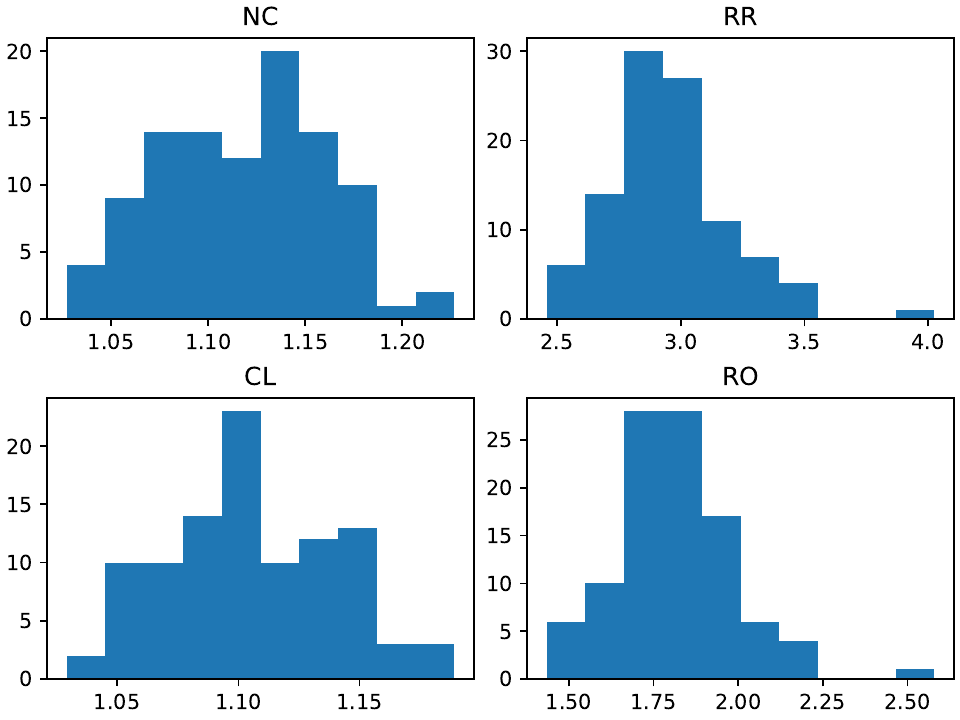}
         \caption{$\eta=0.5$}
         \label{fig:P32-N16-Normal_C0.5}
     \end{subfigure}
     \hfill
     \begin{subfigure}[b]{0.45\textwidth}
         \centering
         \includegraphics[width=\textwidth]{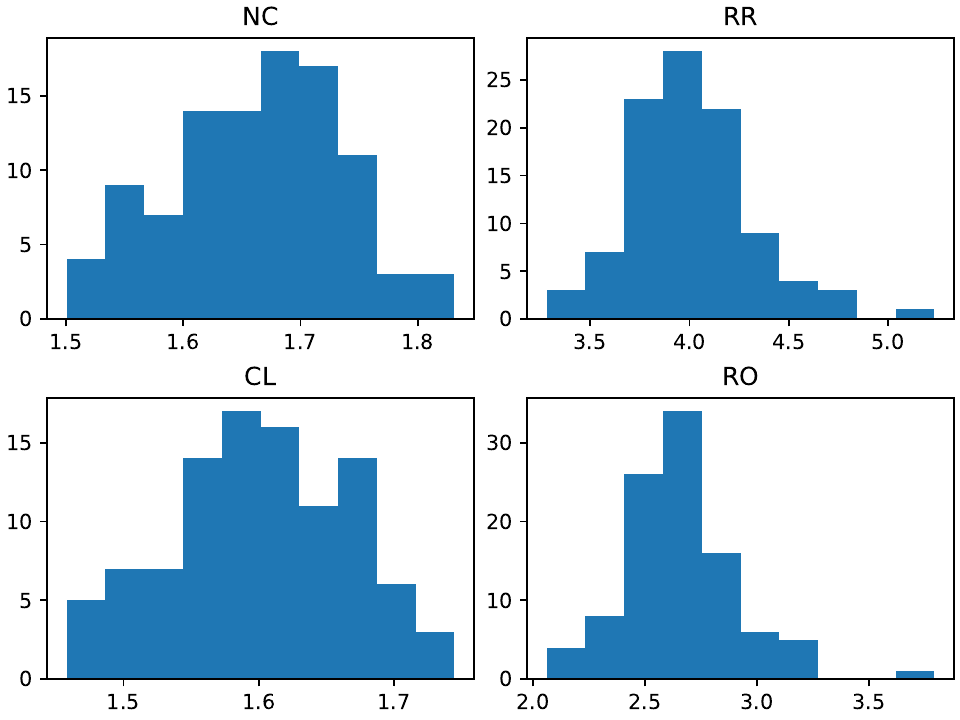}
         \caption{$\eta=2$}
         \label{fig:P32-N16-Normal_C2}
     \end{subfigure}
     \caption{Histograms of the ratios $C^{ALG}/C^{LP}$ for CL, NC, RR and RO for $100$ random problem instances with $L=32$ ports and $n=16$ coflows. Flow sizes follow a Normal distribution with either (a) $\eta=0.5$ or (b) $\eta=2$.\label{fig:P32-N16-Normal}}
 \end{figure}

As for the gamma distribution, UB is quite pessimistic compared to the observed ratios. Regarding $C^{ALG}/C^{LP}$, we observe that when the variability of flow sizes is low the NC performs almost as well as CL and provides almost optimal solutions. We also observe that the variability of the ratio $C^{NC}/C^{LP}$ is quite low in all configurations and that NC provides much better results than RR. These observations are confirmed by the histograms shown in Figure \ref{fig:P32-N16-Normal_C0.5} which report the ratios $C^{ALG}/C^{LP}$ obtained for the $100$ random problem instances in the case $L=32$ and $n=16$. Although the repartition is slightly different, CL and NC provide similar results, which are much better than those obtained with RR or RO.

\begin{table}
\begin{center}
\begin{scriptsize}
\begin{tabular}{l c c cccc cccc cccc}
\toprule
  &  & UB & \multicolumn{4}{c}{CL} & \multicolumn{4}{c}{NC} & \multicolumn{4}{c}{RR} \\
\cmidrule(lr){3-3}\cmidrule(lr){4-7}\cmidrule(lr){8-11}\cmidrule(lr){12-15}
$L$ & $n$ &    & mean  & std & Q1  & Q3  & mean  & std & Q1  & Q3 & mean  & std & Q1  & Q3\\
\midrule
4 & 4 & 7.3 &  1.02  &  0.06  &  0.98  &  1.08  & 1.09  &  0.06  &  1.04  &  1.12  & 1.58  &  0.13  &  1.51  &  1.69 \\
8 & 4 & 8.1 &  1.08  &  0.06  &  1.04  &  1.13  & 1.12  &  0.08  &  1.06  &  1.15  & 1.74  &  0.19  &  1.62  &  1.86 \\
8 & 8 & 8.1 &  1.05  &  0.05  &  1.02  &  1.09  & 1.10  &  0.05  &  1.06  &  1.13  & 2.12  &  0.18  &  2.03  &  2.23 \\
16 & 8 & 8.7 & 1.09  &  0.04  &  1.07  &  1.12  & 1.12  &  0.05  &  1.08  &  1.15  & 2.27  &  0.18  &  2.14  &  2.36 \\
16 & 16 & 8.7 &  1.09  &  0.03  &  1.07  &  1.11  & 1.12  &  0.05  &  1.08  &  1.15  & 2.84  &  0.20  &  2.70  &  2.98 \\
32 & 16 & 9.3 & 1.11  &  0.04  &  1.08  &  1.13  & 1.12  &  0.04  &  1.08  &  1.15  & 2.95  &  0.24  &  2.80  &  3.06 \\
\bottomrule
\end{tabular}
\end{scriptsize}
\caption{$C^{ALG}/C^{LP}$ for flow sizes following a Normal distribution with $\eta=0.5$. \label{tab:LP-Normal_C0.5}}
\end{center}
\end{table}

When the flow size variability increases, we observe, as in the case of the gamma distribution, a slight performance degradation of NC as compared to CL. Nevertheless, the results obtained with NC are much better than those obtained with RR. This can also be observed in the histograms shown in Figure \ref{fig:P32-N16-Normal_C2} for $32$ ports and $16$ coflows. We observe that in this configuration, a random-order policy is much more efficient than RR.

\begin{table}
\begin{center}
\begin{scriptsize}
\begin{tabular}{l c c cccc cccc cccc}
\toprule
  &  & UB & \multicolumn{4}{c}{CL} & \multicolumn{4}{c}{NC} & \multicolumn{4}{c}{RR} \\
\cmidrule(lr){3-3}\cmidrule(lr){4-7}\cmidrule(lr){8-11}\cmidrule(lr){12-15}
$L$ & $n$ &    & mean  & std & Q1  & Q3  & mean  & std & Q1  & Q3 & mean  & std & Q1  & Q3\\
\midrule
4 & 4 & 10.7 & 1.08  &  0.09  &  1.01  &  1.13  & 1.22  &  0.08  &  1.17  &  1.27  & 1.58  &  0.17  &  1.50  &  1.69 \\
8 & 4 &  12.2 & 1.20  &  0.09  &  1.13  &  1.25  & 1.27  &  0.10  &  1.19  &  1.33  & 1.83  &  0.22  &  1.69  &  1.98 \\
8 & 8 &  12.2 & 1.14  &  0.07  &  1.10  &  1.19  & 1.24  &  0.07  &  1.20  &  1.29  & 2.16  &  0.21  &  2.05  &  2.30 \\
16 & 8 &  13.4 & 1.21  &  0.05  &  1.17  &  1.25  & 1.26  &  0.07  &  1.22  &  1.30  & 2.40  &  0.20  &  2.27  &  2.53 \\
16 & 16 &  13.4 & 1.20  &  0.04  &  1.17  &  1.22  & 1.26  &  0.06  &  1.22  &  1.29  & 2.98  &  0.21  &  2.83  &  3.13 \\
32 & 16 &  14.5 & 1.22  &  0.04  &  1.19  &  1.25  & 1.25  &  0.05  &  1.21  &  1.28  & 3.15  &  0.25  &  3.00  &  3.27 \\
\bottomrule
\end{tabular}
\end{scriptsize}
\caption{$C^{ALG}/C^{LP}$ for flow sizes following a Normal distribution with $\eta=1$.\label{tab:LP-Normal_C1}}
\end{center}
\end{table}

\begin{table}
\begin{center}
\begin{scriptsize}
\begin{tabular}{l c c cccc cccc cccc}
\toprule
  &  & UB & \multicolumn{4}{c}{CL} & \multicolumn{4}{c}{NC} & \multicolumn{4}{c}{RR} \\
\cmidrule(lr){3-3}\cmidrule(lr){4-7}\cmidrule(lr){8-11}\cmidrule(lr){12-15}
$L$ & $n$ &    & mean  & std & Q1  & Q3  & mean  & std & Q1  & Q3 & mean  & std & Q1  & Q3\\
\midrule
4 & 4 & 17.3 &  1.37  &  0.14  &  1.25  &  1.46  & 1.66  &  0.13  &  1.57  &  1.75  & 1.92  &  0.24  &  1.77  &  2.09 \\
8 & 4 & 20.3 &  1.57  &  0.15  &  1.47  &  1.69  & 1.72  &  0.16  &  1.63  &  1.82  & 2.30  &  0.31  &  2.12  &  2.52 \\
8 & 8 &  20.3 & 1.48  &  0.11  &  1.41  &  1.56  & 1.68  &  0.11  &  1.61  &  1.76  & 2.65  &  0.29  &  2.45  &  2.87 \\
16 & 8 & 22.8 &  1.60  &  0.08  &  1.55  &  1.66  & 1.71  &  0.10  &  1.65  &  1.78  & 3.07  &  0.27  &  2.91  &  3.26 \\
16 & 16 & 22.8 &  1.57  &  0.07  &  1.53  &  1.62  & 1.70  &  0.09  &  1.64  &  1.76  & 3.74  &  0.26  &  3.56  &  3.91 \\
32 & 16 & 25.1 &  1.60  &  0.06  &  1.55  &  1.65  & 1.66  &  0.07  &  1.61  &  1.71  & 4.01  &  0.31  &  3.81  &  4.17 \\
\bottomrule
\end{tabular}
\end{scriptsize}
\caption{$C^{ALG}/C^{LP}$ for flow sizes following a Normal distribution with $\eta=2$.\label{tab:LP-Normal_C2}}
\end{center}
\end{table}

\begin{figure}
     \centering
	\includegraphics[width=0.6\textwidth]{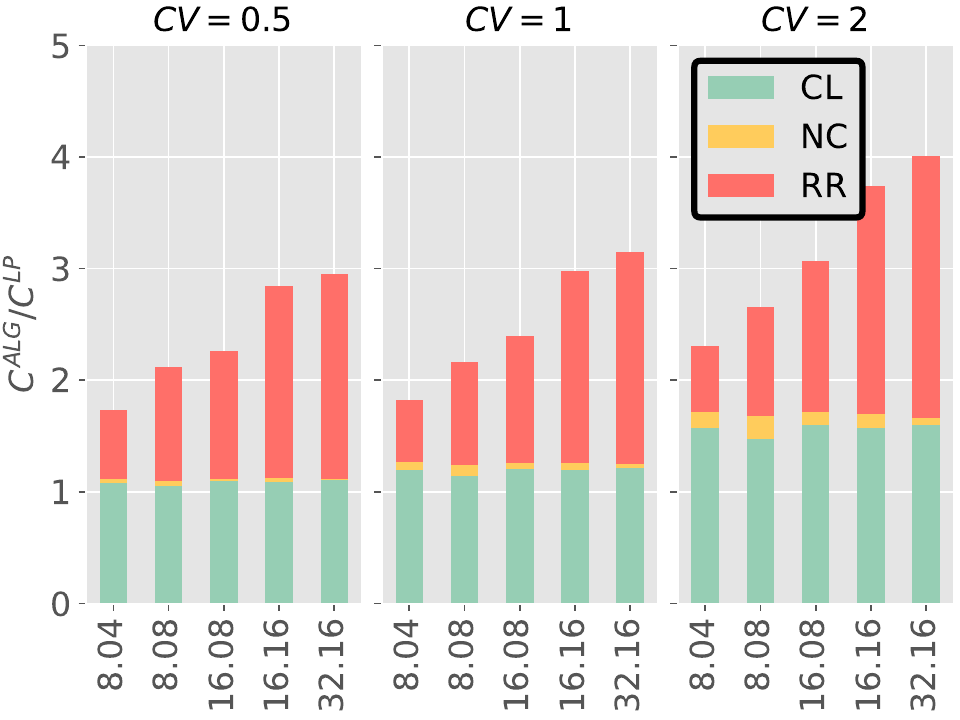}
     \caption{Average values of $C^{ALG}/C^{LP}$ for flow sizes following a Normal distribution with $\eta=0.5$ (left), $\eta=1$ (middle) or $\eta=2$ (right). The number of ports $L$ and the number $n$ of coflows are indicated on the $x$-axis as $L.n$.\label{fig:gauss-means}}
 \end{figure}

Figure \ref{fig:gauss-means} shows the average values (over $100$ random instances) obtained for CL, NC and RR.  The main observations are similar to those made for a gamma distribution of flow sizes.

% Pareto distribution
\subsubsection*{Pareto distribution}
\begin{figure}
     \centering
     \begin{subfigure}[b]{0.45\textwidth}
         \centering
         \includegraphics[width=\textwidth]{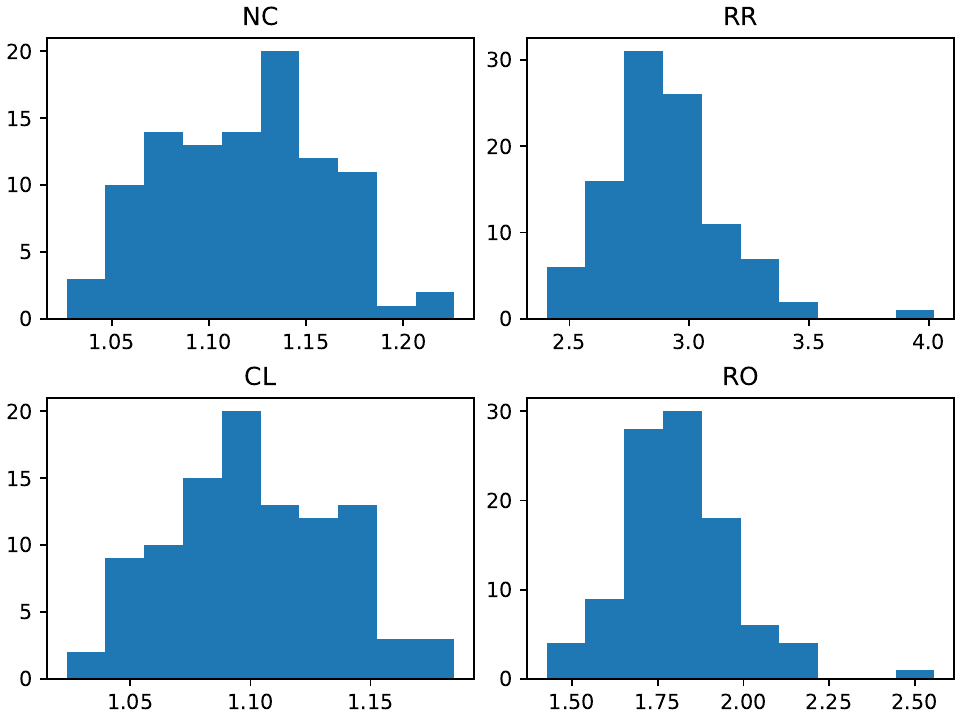}
         \caption{$\eta=0.5$}
         \label{fig:P32-N16-Pareto_C0.5}
     \end{subfigure}
     \hfill
     \begin{subfigure}[b]{0.45\textwidth}
         \centering
         \includegraphics[width=\textwidth]{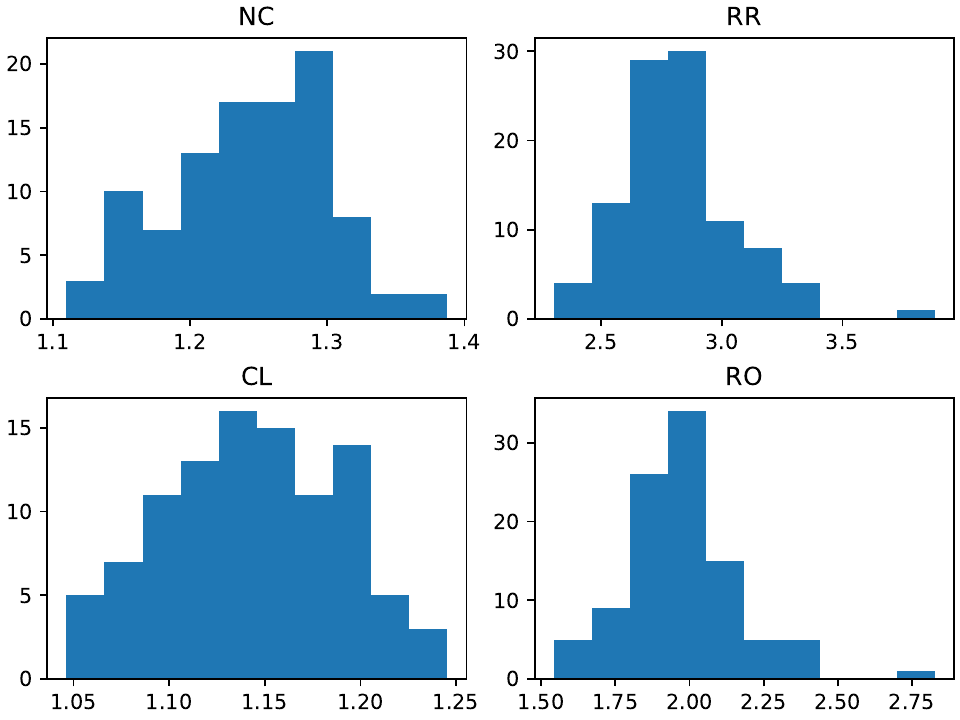}
         \caption{$\eta=2$}
         \label{fig:P32-N16-Pareto_C2}
     \end{subfigure}
     \caption{Histograms of the ratios $C^{ALG}/C^{LP}$ for CL, NC, RR and RO for $100$ random problem instances with $L=32$ ports and $n=16$ coflows. Flow sizes follow a Pareto distribution with either (a) $\eta=0.5$ or (b) $\eta=2$.\label{fig:P32-N16-Pareto}}
 \end{figure}

The statistical results obtained for a Pareto distribution of flow sizes are reported in Tables \ref{tab:LP-Pareto_C0.5}, \ref{tab:LP-Pareto_C1} and \ref{tab:LP-Pareto_C2} for $\eta=0.5$, $\eta=1$ and $\eta=2$, respectively. Figures \ref{fig:P32-N16-Pareto_C0.5} and \ref{fig:P32-N16-Pareto_C2} show the statistical distributions of the ratios $C^{ALG}/C^{LP}$ obtained for $L=32$ ports, $n=16$ coflows and $\eta=0.5$ or $\eta=2$, respectively. Column UB in these tables is the theoretical upper bound on $C^{NC}/C^{LP}$ computed using \eqref{eq:ubpareto}.

As for the gamma and normal distributions, we observe that UB is pessimistic, and that NC is near-optimal and performs almost as well as CL  when the variability of flow sizes is low, and remains quite close to it when the variability of flow sizes increases. In all considered configurations, NC outperforms RR, which in most cases is worst than an random-order policy.

\begin{table}
\begin{center}
\begin{scriptsize}
\begin{tabular}{l c c cccc cccc cccc}
\toprule
  &  & UB & \multicolumn{4}{c}{CL} & \multicolumn{4}{c}{NC} & \multicolumn{4}{c}{RR} \\
\cmidrule(lr){3-3}\cmidrule(lr){4-7}\cmidrule(lr){8-11}\cmidrule(lr){12-15}
$L$ & $n$ &    & mean  & std & Q1  & Q3  & mean  & std & Q1  & Q3 & mean  & std & Q1  & Q3\\
\midrule
4 & 4 & 8.0 &  1.02  &  0.05  &  0.97  &  1.06  & 1.06  &  0.05  &  1.03  &  1.10  & 1.59  &  0.11  &  1.54  &  1.69 \\
8 & 4 & 9.7 &  1.07  &  0.05  &  1.03  &  1.11  & 1.10  &  0.08  &  1.05  &  1.13  & 1.72  &  0.19  &  1.60  &  1.85 \\
8 & 8 & 9.7 &  1.04  &  0.04  &  1.01  &  1.07  & 1.09  &  0.05  &  1.05  &  1.12  & 2.11  &  0.17  &  2.02  &  2.21 \\
16 & 8 & 12.0 &  1.09  &  0.04  &  1.06  &  1.11  & 1.11  &  0.05  &  1.08  &  1.15  & 2.23  &  0.18  &  2.11  &  2.34 \\
16 & 16 & 12.0 & 1.08  &  0.03  &  1.06  &  1.10  & 1.12  &  0.05  &  1.08  &  1.15  & 2.79  &  0.21  &  2.65  &  2.94 \\
32 & 16 &  15.3 & 1.10  &  0.03  &  1.08  &  1.13  & 1.12  &  0.04  &  1.08  &  1.15  & 2.90  &  0.24  &  2.75  &  3.02 \\
\bottomrule
\end{tabular}
\end{scriptsize}
\caption{$C^{ALG}/C^{LP}$ for flow sizes following a Pareto distribution with $\eta=0.5$.\label{tab:LP-Pareto_C0.5}}
\end{center}
\end{table}

\begin{table}
\begin{center}
\begin{scriptsize}
\begin{tabular}{l c c cccc cccc cccc}
\toprule
  &  & UB & \multicolumn{4}{c}{CL} & \multicolumn{4}{c}{NC} & \multicolumn{4}{c}{RR} \\
\cmidrule(lr){3-3}\cmidrule(lr){4-7}\cmidrule(lr){8-11}\cmidrule(lr){12-15}
$L$ & $n$ &    & mean  & std & Q1  & Q3  & mean  & std & Q1  & Q3 & mean  & std & Q1  & Q3\\
\midrule
4 & 4 & 12.0 &  0.99  &  0.05  &  0.94  &  1.03  & 1.08  &  0.06  &  1.03  &  1.11  & 1.52  &  0.11  &  1.47  &  1.62 \\
8 & 4 &  15.3 & 1.08  &  0.06  &  1.03  &  1.13  & 1.14  &  0.09  &  1.08  &  1.19  & 1.69  &  0.20  &  1.56  &  1.82 \\
8 & 8 &  15.3 & 1.04  &  0.05  &  1.01  &  1.08  & 1.13  &  0.06  &  1.10  &  1.18  & 2.03  &  0.17  &  1.96  &  2.13 \\
16 & 8 &  20.0 & 1.10  &  0.04  &  1.08  &  1.13  & 1.17  &  0.06  &  1.13  &  1.21  & 2.19  &  0.18  &  2.05  &  2.29 \\
16 & 16 & 20.0 &  1.09  &  0.04  &  1.07  &  1.11  & 1.17  &  0.06  &  1.13  &  1.21  & 2.71  &  0.20  &  2.57  &  2.85 \\
32 & 16 & 26.6 & 1.12  &  0.04  &  1.09  &  1.15  & 1.18  &  0.05  &  1.14  &  1.21  & 2.85  &  0.24  &  2.69  &  2.96 \\
\bottomrule
\end{tabular}
\end{scriptsize}
\caption{$C^{ALG}/C^{LP}$ for flow sizes following a Pareto distribution with $\eta=1$.\label{tab:LP-Pareto_C1}}
\end{center}
\end{table}

\begin{table}
\begin{center}
\begin{scriptsize}
\begin{tabular}{l c c cccc cccc cccc}
\toprule
  &  & UB & \multicolumn{4}{c}{CL} & \multicolumn{4}{c}{NC} & \multicolumn{4}{c}{RR} \\
\cmidrule(lr){3-3}\cmidrule(lr){4-7}\cmidrule(lr){8-11}\cmidrule(lr){12-15}
$L$ & $n$ &    & mean  & std & Q1  & Q3  & mean  & std & Q1  & Q3 & mean  & std & Q1  & Q3\\
\midrule
4 & 4 & 20.0 &   0.98  &  0.06  &  0.94  &  1.02  & 1.10  &  0.07  &  1.06  &  1.13  & 1.48  &  0.11  &  1.41  &  1.58 \\
8 & 4 &  26.6 & 1.08  &  0.07  &  1.03  &  1.13  & 1.17  &  0.10  &  1.11  &  1.22  & 1.66  &  0.20  &  1.53  &  1.79 \\
8 & 8 &  26.6 & 1.03  &  0.06  &  1.01  &  1.07  & 1.17  &  0.08  &  1.12  &  1.22  & 1.98  &  0.17  &  1.90  &  2.08 \\
16 & 8 & 36.0 & 1.12  &  0.05  &  1.09  &  1.16  & 1.23  &  0.07  &  1.19  &  1.28  & 2.17  &  0.18  &  2.03  &  2.28 \\
16 & 16 & 36.0 & 1.10  &  0.04  &  1.08  &  1.13  & 1.24  &  0.06  &  1.19  &  1.27  & 2.66  &  0.20  &  2.53  &  2.80 \\
32 & 16 & 49.2 & 1.14  &  0.05  &  1.11  &  1.18  & 1.24  &  0.06  &  1.21  &  1.29  & 2.82  &  0.24  &  2.67  &  2.93 \\
\bottomrule
\end{tabular}
\end{scriptsize}
\caption{$C^{ALG}/C^{LP}$ for flow sizes following a Pareto distribution with $\eta=2$.\label{tab:LP-Pareto_C2}}
\end{center}
\end{table}

\begin{figure}
     \centering
	\includegraphics[width=0.6\textwidth]{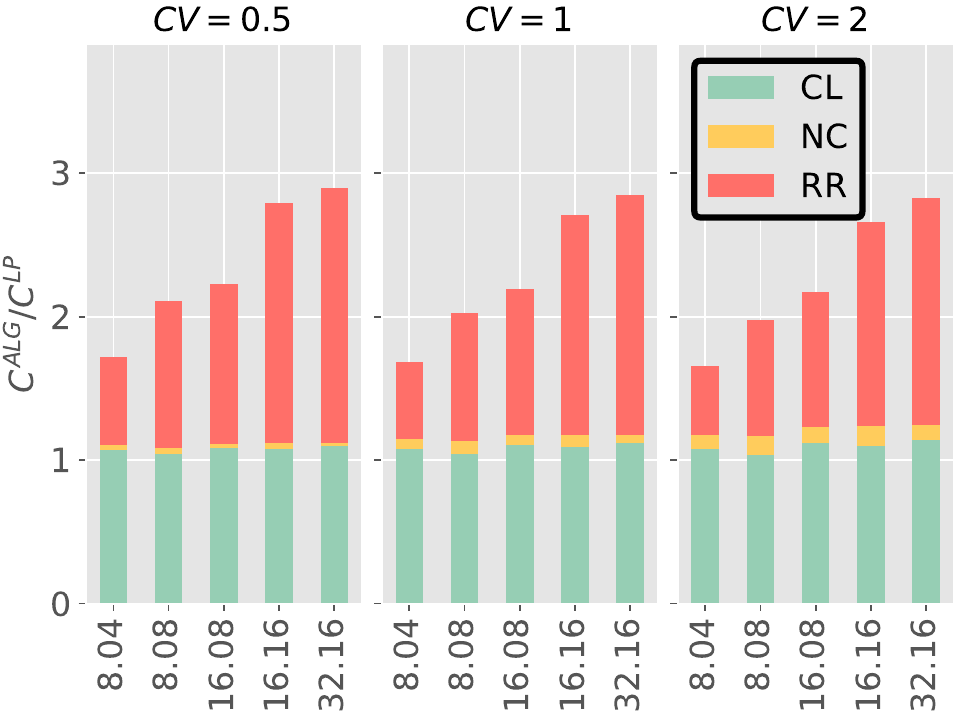}
     \caption{Average values of the ratios $C^{ALG}/C^{LP}$ for flow sizes following a Pareto distribution with $\eta=0.5$ (left), $\eta=1$ (middle) or $\eta=2$ (right). The number of ports $L$ and the number $n$ of coflows are indicated on the $x$-axis as $L.n$.\label{fig:pareto-means}}
 \end{figure}

Figure \ref{fig:pareto-means} shows the average values (over $100$ random instances) obtained for CL, NC and RR.  The main observations are similar to those made for gamma and normal distributions of flow sizes.

% Facebook trace
\subsubsection*{Facebook trace}
\begin{figure}
     \centering
	\includegraphics[width=0.45\textwidth]{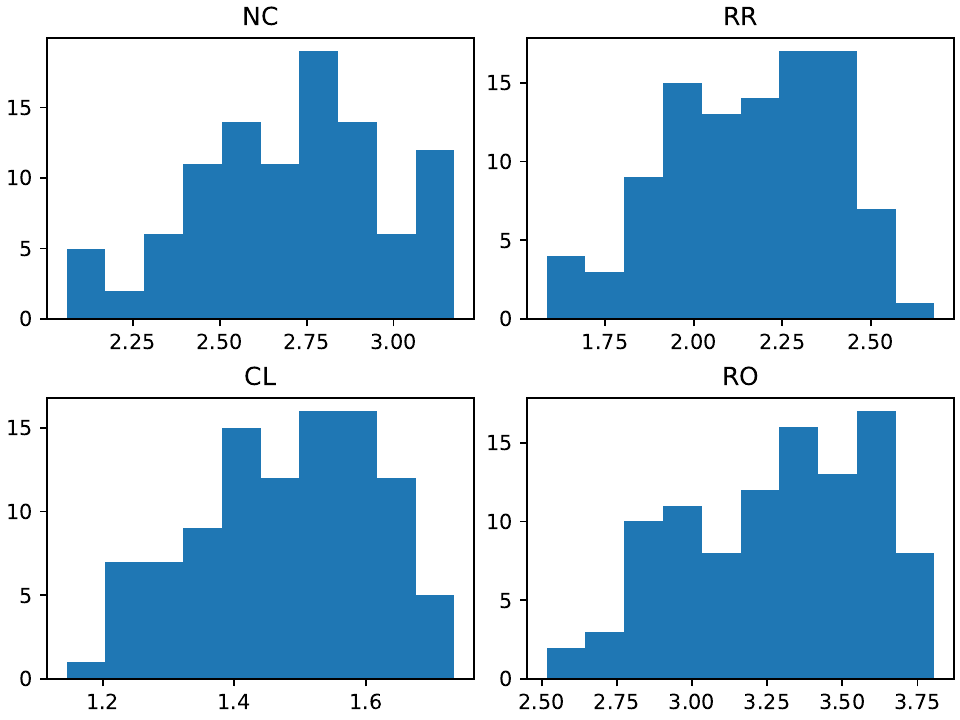}
     \caption{Histograms of the ratios $C^{ALG}/C^{LP}$ for CL, NC, RR and RO for $100$ random problem instances with $L=32$ ports and $n=16$ coflows. Flow sizes are distributed as in the Facebook trace.\label{fig:P32-N16-FB}}
 \end{figure}

\begin{figure}
     \centering
	\includegraphics[width=0.6\textwidth]{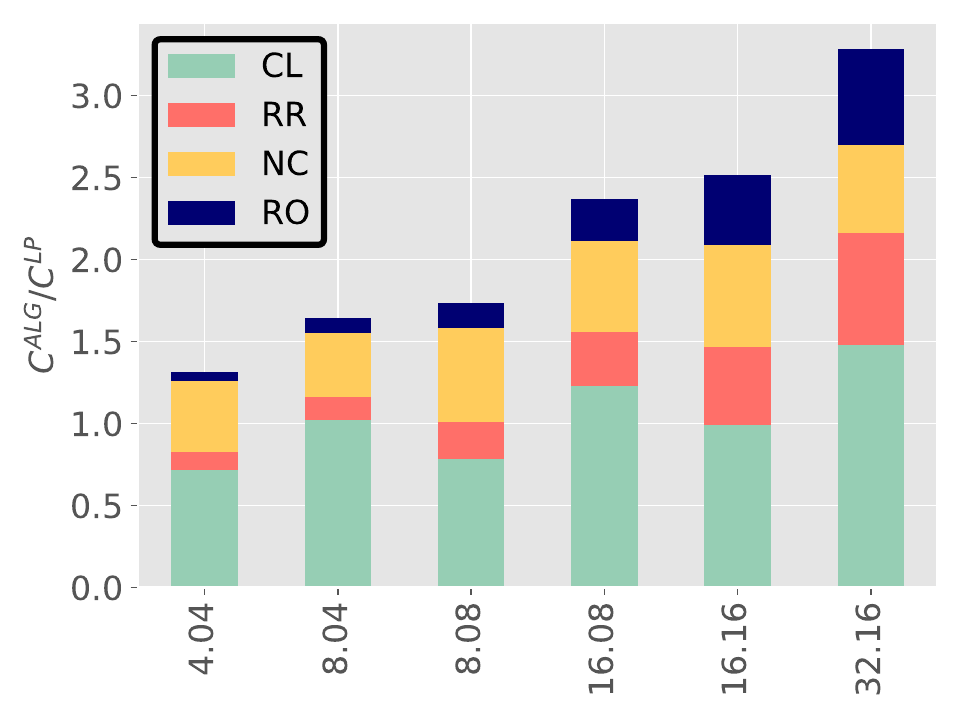}
     \caption{Average values of $C^{ALG}/C^{LP}$ for flow sizes distributed as in the Facebook trace. The number of ports $L$ and the number $n$ of coflows are indicated on the $x$-axis as $L.n$.\label{fig:FB-means}}
 \end{figure}

We now present the results obtained with a flow size distribution derived from the empirical distribution of flow sizes in the Facebook trace. The statistical results obtained in this case are reported in Tables \ref{tab:LP-FB}, and we show the specific distributions of the ratios $C^{ALG}/C^{LP}$ obtained for $L=32$ and $n=16$ in Figure \ref{fig:P32-N16-FB}.Figure \ref{fig:FB-means} shows the average values (over $100$ random instances) obtained for CL, NC and RR.

\begin{table}
\begin{center}
\begin{scriptsize}
\begin{tabular}{l c cccc cccc cccc}
\toprule
  &  & \multicolumn{4}{c}{CL} & \multicolumn{4}{c}{NC} & \multicolumn{4}{c}{RR} \\
\cmidrule(lr){3-6}\cmidrule(lr){7-10}\cmidrule(lr){11-14}
$L$ & $n$    & mean  & std & Q1  & Q3  & mean  & std & Q1  & Q3 & mean  & std & Q1  & Q3\\
\midrule
2 & 2 &  0.71  &  0.02  &  0.70  &  0.73  & 1.00  &  0.02  &  0.99  &  1.02  & 0.76  &  0.02  &  0.75  &  0.77 \\
4 & 4 &  0.72  &  0.10  &  0.65  &  0.79  & 1.26  &  0.17  &  1.14  &  1.36  & 0.83  &  0.12  &  0.73  &  0.90 \\
8 & 4 &  1.02  &  0.19  &  0.90  &  1.13  & 1.55  &  0.30  &  1.36  &  1.74  & 1.16  &  0.23  &  1.03  &  1.31 \\
8 & 8 &  0.79  &  0.11  &  0.72  &  0.85  & 1.58  &  0.22  &  1.43  &  1.74  & 1.01  &  0.17  &  0.91  &  1.11 \\
16 & 8 &  1.23  &  0.17  &  1.14  &  1.34  & 2.11  &  0.34  &  1.87  &  2.36  & 1.56  &  0.25  &  1.38  &  1.72 \\
16 & 16 &  0.99  &  0.10  &  0.93  &  1.06  & 2.09  &  0.19  &  1.95  &  2.23  & 1.47  &  0.17  &  1.36  &  1.61 \\
32 & 16 &  1.48  &  0.14  &  1.39  &  1.58  & 2.70  &  0.27  &  2.51  &  2.90  & 2.16  &  0.24  &  1.95  &  2.35 \\
\bottomrule
\end{tabular}
\end{scriptsize}
\caption{$C^{ALG}/C^{LP}$ for flow sizes distributed as in the Facebook trace.\label{tab:LP-FB}}
\end{center}
\end{table}

Those results are completely different from those obtained for synthetic distribution of flow sizes. For all considered configurations, NC performs poorly compared to CL, and is even significantly worse than RR. The ratio $C^{NC}/C^{LP}$ is also much more variable than for synthetic distributions. For small values of $L$ and $n$, both CL and RR provide average CCT which are lower than $C^{LP}$. We remark that RR can perform better than the lower bound of LP since it falls outside the class of order-based policies. The LP bound is valid only for policies that are within this class. 

In all cases, RR performs much better than NC, which is not much more better than a random order scheduling. As is apparent from Figure \ref{fig:P32-N16-FB}, the performance of NC is similar to that of a random-order policy. This can be explained by the extreme variability of the flow size distribution in the Facebook trace which has a coefficient of variation of $3.27$ for the aggregated flow-sizes of a coflow at a port. While this quantity is not exactly the coefficient of variation of the flow-sizes, we expect these two to be related. If the aggregate coefficient of variation is high, we except that of the flow-sizes to be high as well.

In conclusion, for small instances and low variability of flow-sizes, the proposed NC policy is close to optimal and is better than RR. However, for high variability of flow-sizes, RR outperforms NC and can even be better than the lower bound obtained from LP.

\subsection{Large instances}
\label{subsec:results-large-instances}
We now consider larger problem instances for which the lower bound $C^{LP}$ cannot be computed. As the clairvoyant offline optimum is no more available,
we choose as reference CL, and present the ratios $C^{ALG}/C^{CL}$ obtained with three different algorithms: the non-clairvoyant Sincronia-based priority policy (NC), the round-robin rate allocation (RR ), and \philae\ which is a closely related non-clairvoyant scheduling approach proposed by A. Jajoo \emph{et al.} in \cite{JHL2022} (see also \cite{JHLD2023}). Instead of assuming known probability distributions of flow sizes inferred from historical data, \philae \ pre-schedules sampled flows (called pilot flows) of each coflow and uses their sizes to estimate the average flow size of the coflow. Learned coflow sizes are then used to transmit flows according to a priority-based policy. This pilot-flow-based scheduling method was shown to be empirically effective in \cite{JHL2022}.

As there is no publicly available implementation of \philae, we implemented our own version. In this simplified version, it is assumed that mean flow sizes are known by the coflow scheduler, so there is no need to sample pilot flows and no estimation error. Exact mean flow sizes are used to assign coflows to priority queues according to a contention-and-length-based policy (see policy D in \cite{JHL2022}). Our implementation uses the same default parameters as those used for \aalo \ and \philae\  in \cite{JHL2022} (that is, there are $K=10$ priority queues, with $Q_0^{hi}=10$ MB and $E=10$). As in in \cite{JHL2022}, the bandwidth of a port is shared between the priority queues according to a weighted sharing mechanism and the weights assigned to individual priority queues decrease exponentially by a factor of $10$.

 As before, we average over $1000$ random traffic realizations for each problem instance. We present below the results obtained for a gamma distribution of flow sizes, as well as for flow sizes distributed as in the Facebook trace. The theoretical upper bound (UB) is not shown here since it was already quite pessimistic for small instances.

% Gamma distribution
\subsubsection*{Gamma distribution}
We consider three different coefficients of variations: $\eta=0.2$, $\eta=1$, and $\eta=5$. The mean, standard deviation as well as the first and third quartiles of the ratios $C^{ALG}/C^{CL}$ are reported in Table \ref{tab:LARGE-Gamma_C0.2} (resp. Table~\ref{tab:LARGE-Gamma_C1} and Table~\ref{tab:LARGE-Gamma_C5}) for flow sizes following a gamma distribution with $\eta = 0.2$ (resp. $\eta=1$ and $\eta=5$). For $\eta=0.2$ and $\eta=1$, we observe that for all considered configurations of the numbers of ports and coflows, NC performs almost as well as CL on average, and that the variability of the ratio $C^{NC}/C^{CL}$ is quite low.  We also observe that the ratio $C^{RR}/C^{CL}$ has a much higher mean and standard deviation, which shows that RR performs poorly as compared to our NC policy when the variability of flow sizes is low. It is also interesting to note that the results obtained with \philae\ are better than those with RR but worse than that of NC. The weighted scheme of \philae\ thus outperforms the unweighted RR for small values of $\eta$.

\begin{table*}
	\begin{center}
		\caption{$C^{ALG}/C^{CL}$ for flow sizes following a Gamma distribution with $\eta=0.2$ and large instances. 
			\label{tab:LARGE-Gamma_C0.2}}
		\begin{scriptsize}
			\begin{tabular}{l c cccc cccc cccc}
				\toprule
				%&  &  \multicolumn{8}{c}{$\eta=0.2$} & \multicolumn{8}{c}{$\eta=5$} \\
				%\cmidrule(lr){3-10}\cmidrule(lr){11-14}
				&  &  \multicolumn{4}{c}{NC} & \multicolumn{4}{c}{RR} & \multicolumn{4}{c}{PH} \\
				\cmidrule(lr){3-6}\cmidrule(lr){7-10}\cmidrule(lr){11-14}
				$L$ & $n$  & mean  & std & Q1  & Q3 & mean  & std & Q1  & Q3 & mean  & std & Q1  & Q3 \\
				\midrule
				20 & 10 &  1.01  &  0.02  &  1.00  &  1.01  & 2.39  &  0.25  &  2.24  &  2.53 & 2.22  &  0.31  &  2.03  &  2.45\\
				20 & 20 &  1.01  &  0.02  &  1.00  &  1.01  & 2.86  &  0.22  &  2.72  &  2.98 & 1.98  &  0.16  &  1.87  &  2.09\\
				20 & 30 &  1.01  &  0.02  &  1.00  &  1.01  & 3.21  &  0.22  &  3.06  &  3.38 & 2.41  &  0.18  &  2.29  &  2.52 \\
				20 & 40 &  1.01  &  0.01  &  1.00  &  1.02  & 3.38  &  0.21  &  3.22  &  3.55 & 2.74  &  0.18  &  2.63  &  2.87 \\
				20 & 60 &  1.01  &  0.01  &  1.00  &  1.02  & 3.57  &  0.19  &  3.44  &  3.69 & 3.14  &  0.14  &  3.05  &  3.23\\
				40 & 20 &  1.00  &  0.01  &  1.00  &  1.01  & 2.95  &  0.28  &  2.78  &  3.09 & 2.33  &  0.19  &  2.19  &  2.46\\
				40 & 40 &  1.01  &  0.01  &  1.00  &  1.02  & 3.44  &  0.23  &  3.28  &  3.59 &  3.08  &  0.18  &  2.94  &  3.20\\
				40 & 60 &  1.01  &  0.01  &  1.00  &  1.02  & 3.97  &  0.27  &  3.81  &  4.15 &  2.92  &  0.44  &  2.56  &  3.30\\
				40 & 80 &  1.01  &  0.01  &  1.00  &  1.02  & 3.83  &  0.22  &  3.67  &  3.96 &  2.30  &  0.27  &  2.11  &  2.42 \\
				\bottomrule
			\end{tabular}
		\end{scriptsize}
	\end{center}
\end{table*}

\begin{table*}
	\caption{$C^{ALG}/C^{CL}$ for flow sizes following a Gamma distribution with $\eta=1$ and large instances. \label{tab:LARGE-Gamma_C1}}
	\begin{center}
		\begin{scriptsize}
			\begin{tabular}{l c cccc cccc cccc cccc}
				\toprule
				&  &  \multicolumn{4}{c}{NC} & \multicolumn{4}{c}{RR} & \multicolumn{4}{c}{PH}\\
				\cmidrule(lr){3-6}\cmidrule(lr){7-10}  \cmidrule(lr){11-14}
				$L$ & $n$  & mean  & std & Q1  & Q3 & mean  & std & Q1  & Q3 & mean  & std & Q1  & Q3\\
				\midrule
				20 & 10 &  1.06  &  0.04  &  1.03  &  1.08  & 2.12  &  0.16  &  2.04  &  2.22 & 2.00  &  0.22  &  1.87  &  2.15 \\
				20 & 20 &  1.07  &  0.03  &  1.04  &  1.08  & 2.56  &  0.14  &  2.47  &  2.65 & 1.88  &  0.14  &  1.78  &  1.98 \\
				20 & 30 &  1.07  &  0.02  &  1.06  &  1.09  & 2.86  &  0.15  &  2.77  &  2.97 & 2.28  &  0.16  &  2.20  &  2.41 \\
				20 & 40 &  1.08  &  0.02  &  1.06  &  1.09  & 3.02  &  0.14  &  2.93  &  3.13 & 2.59  &  0.15  &  2.51  &  2.69 \\
				20 & 60 &  1.08  &  0.02  &  1.06  &  1.09  & 3.21  &  0.12  &  3.13  &  3.30 & 2.94  &  0.10  &  2.88  &  3.01 \\
				40 & 20 &  1.04  &  0.03  &  1.02  &  1.05  & 2.72  &  0.20  &  2.60  &  2.83 & 2.24  &  0.18  &  2.11  &  2.35 \\
				40 & 40 &  1.05  &  0.02  &  1.03  &  1.06  & 3.18  &  0.18  &  3.05  &  3.32 & 2.93  &  0.15  &  2.82  &  3.05 \\
				40 & 60 &  1.05  &  0.02  &  1.04  &  1.06  & 3.40  &  0.18  &  3.29  &  3.52 & 2.79  &  0.40  &  2.46  &  3.14 \\
				40 & 80 &  1.05  &  0.02  &  1.04  &  1.06  & 3.58  &  0.17  &  3.45  &  3.70 & 2.20  &  0.23  &  2.05  &  2.28 \\
				\bottomrule
			\end{tabular}
		\end{scriptsize}
	\end{center}
\end{table*}

\begin{table*}
	\begin{center}
		\caption{$C^{ALG}/C^{CL}$ for flow sizes following a Gamma distribution with $\eta=5$ and large instances. 
			\label{tab:LARGE-Gamma_C5}}
		\begin{scriptsize}
			\begin{tabular}{l c cccc cccc cccc}
				\toprule
				&  &  \multicolumn{4}{c}{NC} & \multicolumn{4}{c}{RR} & \multicolumn{4}{c}{PH}  \\
				\cmidrule(lr){3-6}\cmidrule(lr){7-10}\cmidrule(lr){11-14}
				$L$ & $n$  & mean  & std & Q1  & Q3 & mean  & std & Q1  & Q3 & mean  & std & Q1  & Q3\\
				\midrule
				20 & 10 &  1.47  &  0.12  &  1.39  &  1.55  & 1.37  &  0.06  &  1.34  &  1.41  &  1.36  &  0.05  &  1.33  &  1.39\\
				20 & 20 &  1.62  &  0.10  &  1.55  &  1.69  & 1.59  &  0.06  &  1.56  &  1.64 & 1.56  &  0.07  &  1.51  &  1.62\\
				20 & 30 &  1.70  &  0.10  &  1.62  &  1.74  & 1.73  &  0.06  &  1.70  &  1.77 & 1.72  &  0.05  &  1.70  &  1.75\\
				20 & 40 &  1.73  &  0.09  &  1.66  &  1.78  & 1.84  &  0.04  &  1.81  &  1.87 & 1.84  &  0.04  &  1.81  &  1.86\\
				20 & 60 &  1.76  &  0.09  &  1.70  &  1.82  & 1.98  &  0.04  &  1.95  &  2.01 & 1.99  &  0.04  &  1.96  &  2.01\\
				40 & 20 &  1.31  &  0.06  &  1.26  &  1.35  & 1.78  &  0.06  &  1.75  &  1.83 & 1.71  &  0.09  &  1.64  &  1.78\\
				40 & 40 &  1.35  &  0.05  &  1.31  &  1.39  & 2.09  &  0.06  &  2.04  &  2.13 & 2.06  &  0.07  &  2.02  &  2.11\\
				40 & 60 &  1.36  &  0.04  &  1.33  &  1.38  & 2.28  &  0.05  &  2.25  &  2.32 & 2.04  &  0.16  &  1.92  &  2.18\\
				40 & 80 &  1.36  &  0.03  &  1.34  &  1.39  & 2.41  &  0.04  &  2.38  &  2.44 & 1.81  &  0.06  &  1.77  &  1.82\\
				\bottomrule
			\end{tabular}
		\end{scriptsize}
	\end{center}
\end{table*}

On the other hand, for $\eta=5$ (that is, flow sizes with high variability), a switch with $20$ ports and $10$ or $20$ coflows, the average performance of RR is better than but close to that of NC. However, for a higher number of coflows or larger switches with $40$ ports, RR becomes worse than NC. This observation is similar to that for small instances and $\eta=2$ (see Table~\ref{tab:LP-Gamma_C2}) for which RR was better for smaller instances but worse for larger ones. The performance of \philae\ is close to that of RR. This is probably due to default configuration that we used for priority queues and for which many coflows fall in the same queue.

\begin{figure}
\begin{subfigure}[b]{0.45\textwidth}
	\centering
\includegraphics[width=\textwidth]{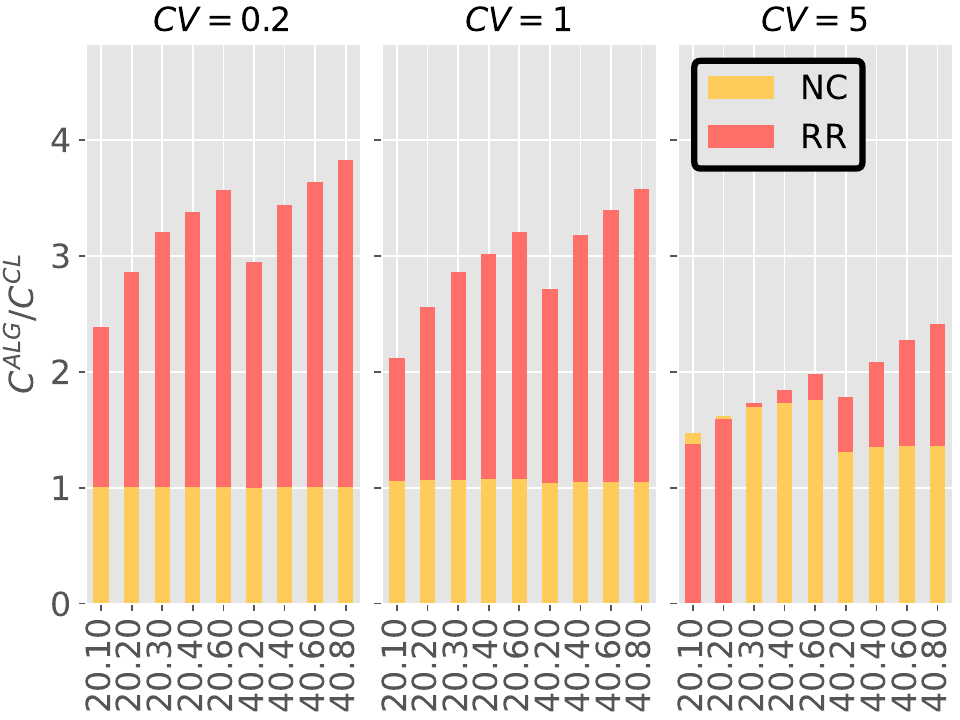}
\caption{$C^{NC}/C^{CL}$ and $C^{RR}/C^{CL}$ with $\eta=0.2$ (left), $\eta=1$ (middle) or $\eta=5$ (right). \label{fig:gamma-means-large-rr}}
\end{subfigure}
\hspace{3mm}
\begin{subfigure}[b]{0.45\textwidth}
	\centering
	\includegraphics[width=\textwidth]{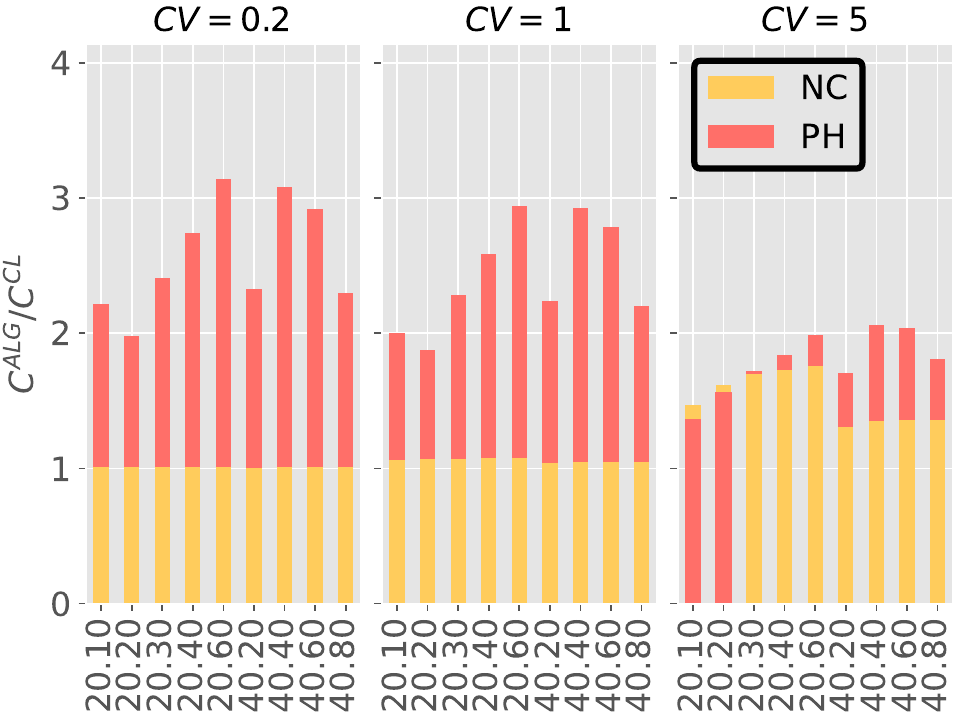}
	\caption{$C^{NC}/C^{CL}$ and $C^{PH}/C^{CL}$ with $\eta=0.2$ (left), $\eta=1$ (middle) or $\eta=5$ (right). \label{fig:gamma-means-large-philae}}
\end{subfigure}
\caption{Average values of the ratios $C^{ALG}/C^{CL}$ for gamma distribution of flow sizes with different coefficient of variations. Large instances. The numbers of ports $L$ and of coflows $n$ are indicated on the $x$-axis as $L.n$.}
\label{fig:gamma-means-large}
\end{figure}

Figure \ref{fig:gamma-means-large} provides a synthetic view of the above results. It shows the average values (over $100$ random instances) of $C^{ALG}/C^{CL}$ obtained for NC and RR. 

% Facebook trace
\subsubsection*{Facebook trace}
\begin{figure}
\centering
\includegraphics[width=0.4\textwidth]{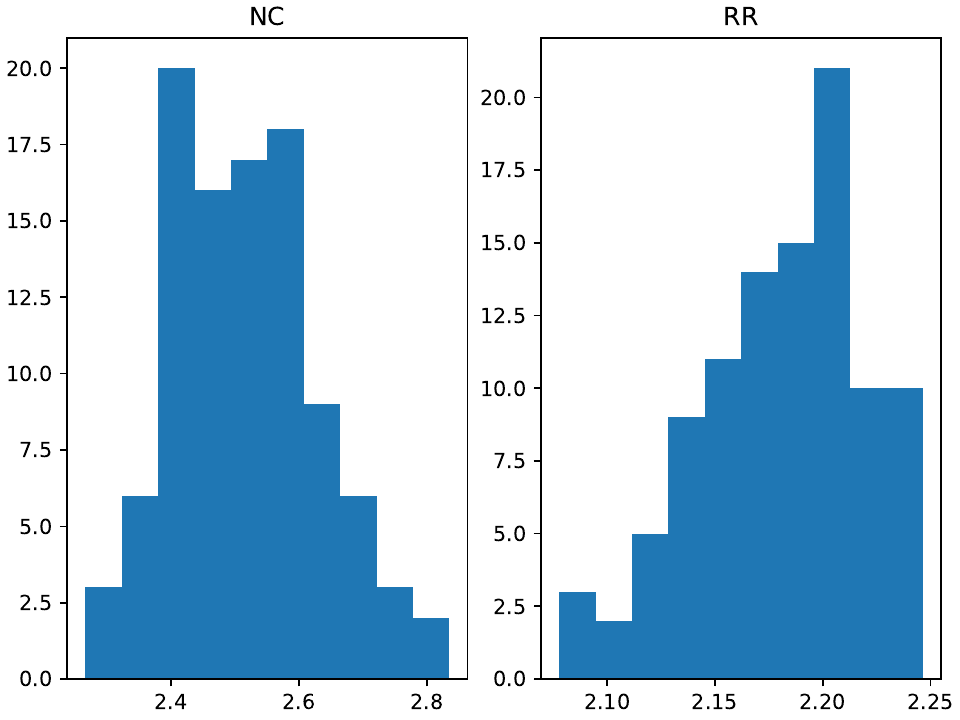}
\caption{Histograms of the ratios $C^{ALG}/C^{CL}$ for $100$ random problem instances with $L=40$ ports and $n=80$ coflows. Flow sizes are distributed as in the Facebook trace.\label{fig:P40-N80-FB-PH}}
\end{figure}

The statistical results for flow size distribution derived from the empirical distribution of flow sizes in the Facebook trace is reported in Table~\ref{tab:LARGE-FB}, and we show the specific distributions of the ratios $C^{ALG}/C^{CL}$ obtained for {$L=40$ and $n=80$} in Figure~\ref{fig:P40-N80-FB-PH}.

For instances with $20$ ports, NC performs poorly compared to CL, and is even significantly worse than RR. However, we observe that NC gets closer to RR for larger instances with $40$ ports. This observation is similar to the one for small instances (see Table~\ref{tab:LP-FB}): NC gets closer RR in performance as the instances become larger.  We believe that RR is better for smaller instances because of high variability of flow sizes. An intuitive explanation for this phenomenon is as follows: as the number of flows increases, the aggregate coefficient of variation at a port will decrease thus favoring NC over RR. The performance of Philae is marginally worse than that of RR. Since there are several parameters to be set in Philae, it is possible that the performance will depend on the specific choice of the parameters. The results presented here are for the values mentioned previously.                                                                                                                                                                               

One avenue for further research would be to conceive adaptive algorithms that choose between RR and NC based on the variability of the flow sizes.

\begin{table*}
	\caption{$C^{ALG}/C^{CL}$ for flow sizes distributed as in the Facebook trace and large instances. \label{tab:LARGE-FB}}
	\begin{center}
		\begin{scriptsize}
			\begin{tabular}{l c cccc cccc cccc cccc}
				\toprule
				&  &  \multicolumn{4}{c}{NC} & \multicolumn{4}{c}{PH} & \multicolumn{4}{c}{RR} \\
				\cmidrule(lr){3-6}\cmidrule(lr){7-10}\cmidrule(lr){11-14}
				$L$ & $n$  & mean  & std & Q1  & Q3 & mean  & std & Q1  & Q3 & mean  & std & Q1  & Q3\\
				\midrule
				20 & 10 & 1.83  &  0.13  &  1.74  &  1.90  & 1.34  &  0.10  &  1.29  &  1.36 & 1.32  &  0.06  &  1.28  &  1.36 \\
				20 & 20 & 2.32  &  0.15  &  2.21  &  2.38  & 1.82  &  0.11  &  1.78  &  1.87 & 1.56  &  0.06  &  1.53  &  1.60\\
				20 & 30 & 2.54  &  0.14  &  2.43  &  2.62  & 2.00  &  0.06  &  1.97  &  2.05 & 1.72  &  0.05  &  1.69  &  1.76\\
				20 & 40 & 2.71  &  0.13  &  2.59  &  2.78  & 2.11  &  0.07  &  2.06  &  2.15 & 1.86  &  0.05  &  1.83  &  1.89\\
				20 & 60 & 2.90  &  0.15  &  2.78  &  3.02  & 2.24  &  0.07  &  2.19  &  2.28 & 2.03  &  0.05  &  2.00  &  2.07\\
				40 & 20 & 1.99  &  0.14  &  1.89  &  2.04  & 1.67  &  0.07  &  1.62  &  1.71 & 1.54  &  0.05  &  1.51  &  1.57\\
				40 & 40 & 2.30  &  0.14  &  2.20  &  2.37  & 1.91  &  0.05  &  1.88  &  1.94 & 1.84  &  0.04  &  1.81  &  1.87\\
				40 & 60 & 2.44  &  0.11  &  2.35  &  2.49  & 2.13  &  0.06  &  2.10  &  2.17 & 2.04  &  0.04  &  2.02  &  2.07\\
				40 & 80 & 2.52  &  0.12  &  2.43  &  2.59  & 2.34  &  0.07  &  2.29  &  2.39 & 2.18  &  0.04  &  2.16  &  2.21\\
				\bottomrule
			\end{tabular}
		\end{scriptsize}
		\color{black}
	\end{center}
\end{table*}

% Numerical Results
\section{Concluding remarks}
\label{sec:conclusion}
We addressed the coflow scheduling problem in the stochastic setting, assuming that the objective is to minimize the weighted expected completion time of coflows. We proposed a simple priority policy in which the priority order is computed with \sincronia using the expected processing times instead of the unknown actual processing times. Performance guarantees are established for this policy using a polyhedral relaxation of the performance space of stochastic coflow scheduling policies, first for arbitrary probability distributions of processing times, and then for some specific distributions. Empirical results show that this priority policy performs very well in practice, and much better than the theoretical upper bounds suggest.

We see the following possible future directions emanating from this work. First, we plan to design preemptive scheduling policies leveraging the information on the attained service of coflows.
Another interesting direction will be to propose policies that choose between RR and NC based on the variability of the flow-sizes.
Further, it will also be interesting to compute the approximation ratio for a richer class of policies than the order-based ones in this paper. As we saw, certain policies like RR are not included in this class. Finally, we intend to establish tighter bounds for some other specific distributions of processing times, and hope to propose methods for computing the mean CCT in some asymptotic regimes such as large number of coflows and ports.

\bibliography{references}

\appendix
\section{Proof of Theorem \ref{thm:stoch-par-inequalities}}
\label{app:proof-stoch-par-inequalities}
\begin{proof}
Let $r$ be a work-conserving and nonanticipative rate-allocation policy and define the random variable $Z_{\ell,k} = C_k(r)  - P_{\ell,k}$. Consider any fixed realization of flow volumes. The latter yields a specific realization of the processing times $p_{\ell,k}$ and of the completion times $c_k(r)$ (we use lowercase letters to denote the particular values that our random variables take for this specific realization of flow volumes). Assume that, for this realization, coflows complete in the order $\pi$, that is, $c_{\pi(1)}(r) \leq c_{\pi(2)}(r) \leq \ldots \leq c_{\pi(n)}(r)$. As coflow $\pi(k)$ completes after all coflows $\pi(j)$, $j=1,\ldots,k-1$, we clearly have $c_{\pi(k)}(r) \geq \sum_{j \leq k} p_{\ell,\pi(j)}$ for all ports $\ell$.  It yields $z_{\ell,\pi(k)} \geq \sum_{j<k} p_{\ell,\pi(j)} \delta_{\pi(j)}^A$ for any subset $A \subseteq \cC$ , where $\delta_{\pi(j)}^A$ is $1$ if $\pi(j) \in A$ and $0$ otherwise. We thus obtain

\begin{align}
\sum_{j \in A} p_{\ell,j} z_{\ell,j} & =  \sum_{k} p_{\ell,\pi(k)} z_{\ell,\pi(k)} \delta_{\pi(k)}^A, \nonumber \\
& \geq  \sum_{k}  \sum_{j<k} p_{\ell,\pi(k)} \delta_{\pi(k)}^A p_{\ell,\pi(j)} \delta_{\pi(j)}^A, \nonumber  \\
& = \frac{1}{2} \sum_{j, k:  j \neq k} p_{\ell,\pi(k)} \delta_{\pi(k)}^A p_{\ell,\pi(j)} \delta_{\pi(j)}^A,  \nonumber  \\
& =  \frac{1}{2} \sum_{j, k \in A, j \neq k} p_{\ell,k} p_{\ell,j},  \label{eq:ineq-realization}
\end{align}

\noindent and the lower bound is independent of the considered order. Conditioning on the order $\pi$ and taking expectation on both sides of \eqref{eq:ineq-realization}, it follows that

\begin{align*}
 \E \left [ \sum_{j \in A}  P_{\ell,j} Z_{\ell,j} \ | \ \mbox{order } \pi \right ] & \geq \frac{1}{2} \sum_{j, k \in A, j \neq k} \E \left [ P_{\ell,k}  P_{\ell,j} \right ], \\
& = \frac{1}{2} \sum_{j, k \in A, j \neq k}  \mu_{\ell,k}  \mu_{\ell,j}, \\
& =  \frac{1}{2} \left ( \sum_{k \in A} \mu_{\ell,k} \right )^2  - \frac{1}{2} \sum_{k \in A} \mu_{\ell,k}^2,
\end{align*}

\noindent where the first equality follows from the independence of processing times. In general, for an arbitrary rate-allocation policy $r$, several completion orders are possible. However, if we assume that $r$ is an order-based policy, the order $\pi$ in which coflows complete is unique and depends only on the partial order implemented by this policy, and not on the actual realization of the processing times. Moreover, as processing times are independent, the random variables $P_{\ell,k}$ and $Z_{\ell,k}$ are statistically independent in this case, so that $\E \left [ P_{\ell,k} Z_{\ell,k} \right ] = \mu_{\ell,k} \E \left [ Z_{\ell,k} \right ]$. Hence

\[
 \E \left [ \sum_{j \in A}  P_{\ell,j} Z_{\ell,j} \ | \ \mbox{order } \pi \right ]  = \E \left [ \sum_{j \in A} P_{\ell,j}  Z_{\ell,j} \right ] = \sum_{j \in A} \mu_{\ell,j}  \E \left [ Z_{\ell,j} \right ],
\]

\noindent so that

\[
\sum_{j \in A} \mu_{\ell,j}  \E \left [ Z_{\ell,j} \right ] \geq \frac{1}{2} \left ( \sum_{k \in A} \mu_{\ell,k} \right )^2  - \frac{1}{2} \sum_{k \in A} \mu_{\ell,k}^2,
\]

	\noindent and we can conclude the proof with $\E \left [ Z_{\ell,j} \right ]= \E \left [C_j(r) \right ] - \mu_{\ell,j}$.
\end{proof}

\section{Proof of Lemma~\ref{lem:bound-on-bottleneck-gamma}}
\label{pr:lem3}
The proof follows the standard approach of using the moment generating function. We give the details here since a similar proof is also used for Gaussian processing times.
\begin{proof}
	First, observe that 
	\begin{equation}
		\max_{\ell,j} \beta_{\ell,j} = \max_{\ell,j} \frac{\sigma^2_{\ell,j}}{\mu_{\ell,j}} \leq \gamma \mu_{min}.\label{eq:22}
	\end{equation}
	
	Let $X_{\ell,k} = \sum_{j=1}^k P_{\ell,j}$ and $Y_k=\max_\ell \frac{X_{\ell,k}}{\sum_i\mu_{\ell,i}}$. Using Jensen's inequality we obtain 
	\begin{align}
		e^{t \E [Y_k]} &\leq \E \left [ e^{tY_k} \right ] = \E \left [ \max_\ell e^{t\frac{X_{\ell,k}}{\sum_i\mu_{\ell,i}}} \right ] \leq \sum_\ell \E \left [ e^{t \frac{X_{\ell,k}}{\sum_i\mu_{\ell,i}}} \right ] \nonumber\\
		&= \sum_\ell \prod_{j=1}^k \E \left [ e^{\frac{t}{\sum_i\mu_{\ell,i}} P_{\ell,j}} \right ] \label{eq:21}
	\end{align}
	with the last inequality following from the independence of $P_{\ell,j}$. Since $P_{\ell,j} \sim \Gamma(s_{\ell,j},\beta_{\ell,j})$, 
	\begin{align}
		\E[e^{tP_{\ell,j}}] &= (1-\beta_{\ell,j} t)^{-s_{\ell,j}} = (1-\beta_{\ell,j} t)^{-\frac{\mu_{\ell,j}}{\beta_{\ell,j}}}\nonumber \\
		&\leq (1-\gamma\mu_{min} t)^{-\frac{\mu_{\ell,j}}{\gamma\mu_{min}}},\quad t < (\gamma\mu_{min}^{-1}) \label{eq:25},
	\end{align} where the second equality follows from the fact that $s_{\ell,j}\beta_{\ell,j} = \mu_{\ell,j}$ and the last inequality is a consequence of the fact that the RHS is increasing in $\beta_{\ell,j}$ and $\beta_{\ell,j} < \gamma\mu_{min}$ (see \eqref{eq:22}). Substituting \eqref{eq:25} in \eqref{eq:21}, for $t < \beta_{max}^{-1}$,
	\begin{align*}
		\sum_\ell \prod_{j=1}^k \E \left [ e^{\frac{t}{\sum_i\mu_{\ell,i}} P_{\ell,j}} \right ] &\leq \sum_\ell \left ( \frac{1}{1-\frac{\gamma\mu_{min}}{\sum_j\mu_{\ell,j}} t} \right )^{\frac{\sum_j\mu_{\ell,j}}{\gamma\mu_{min}}}\\
		& \leq L\left ( \frac{1}{1-\gamma t} \right )^{\frac{1}{\gamma}},
	\end{align*}
	where the last inequality uses the fact that the RHS is decreasing in $\sum_j\mu_{\ell,j}$ so that we can substitute $\sum_j\mu_{\ell,j}$ by $\mu_{min}$.
	Thus,
	\[
	\E [Y_k] \leq  \frac{\gamma\log(L) - \log(1-\gamma t)}{\gamma t}, \mbox{ for } t < \gamma^{-1}.
	\]
	
	The minimum of the RHS is obtained for 
	\begin{equation}
		\gamma\hat{t}: \frac{1}{1-\gamma \hat{t}} = \frac{\gamma\log(L) - \log(1-\gamma \hat{t})}{\gamma\hat{t}}
		\label{eq:zeroht}
	\end{equation}
	and 
	\[
	\E [Y_k] \leq \frac{1}{1-\gamma \hat{t}}.
	\]
	We now show that the RHS in \eqref{eq:ubgamma} is an upper bound for $\frac{1}{1-\gamma \hat{t}} =:z$. Rearranging \eqref{eq:zeroht}, $z$ is the solution of
	\begin{equation}
		z -\log(z) = 1 + \gamma\log L.
		\label{eq:z}
	\end{equation}
	Note that the LHS is an increasing function of $z$ and any $z$ satisfying the equality is larger than $1$. Applying the bound \cite{T07} 
	$$
	\log(z) \leq \frac{(z-1)(z+1)}{2z}, \,\mbox{for } z \geq 1,
	$$
	in the LHS of \eqref{eq:z} we get
	$$
	z - \frac{(z-1)(z+1)}{2z} \leq z -\log(z) = c,
	$$
	which leads to 
	$$
	z \leq c + \sqrt{c^2 - 1}  \Leftrightarrow \log(z) \leq \log(c + \sqrt{c^2 - 1})
	$$
	Substituting this bound for $\log(z)$ in \eqref{eq:z}, we obtain the stated result.
\end{proof}

% Proof of Lemma~\ref{lem:bound-on-bottleneck}
\section{Proof of Lemma~\ref{lem:bound-on-bottleneck}}
\label{pr:lem:bound-on-bottleneck}
\begin{proof}
To simplify notations, let $X_\ell = \sum_{j=1}^k P_{\ell,j}$ and $Y=\max_\ell X_\ell$. We follow a standard approach by first deriving a bound on the moment-generating function of $Y$:

\begin{align}
exp \left ( t \E [Y] \right ) & \leq \E \left [ exp(tY) \right ], \label{eq:mgf-step1} \\
& = \E \left [ \max_\ell exp(t X_\ell) \right ], \label{eq:mgf-step2} \\
& \leq \sum_\ell \E \left [ exp(t X_\ell) \right ], \label{eq:mgf-step3} \\
& = \sum_\ell \E \left [ \prod_{j=1}^k exp(t P_{\ell,j}) \right ], \label{eq:mgf-step4} \\
& = \sum_\ell \prod_{j=1}^k \E \left [ exp(t P_{\ell,j}) \right ], \label{eq:mgf-step5} 
\end{align}

\noindent where inequality \eqref{eq:mgf-step1} follows from Jensen's inequality and eqality \eqref{eq:mgf-step5} follows from the independence assumption. As we assume a normal distribution for the $P_{\ell,j}$, we have $\E \left [ exp(t P_{\ell,j}) \right ]= exp \left (t \mu_{\ell,j} + \frac{\sigma_{\ell,j}^2}{2} t^2 \right )$, where $\sigma_{\ell,j}^2 = \eta_{\ell,j}^2 \mu_{\ell,j}^2$ is the variance. It yields

\begin{align*}
exp \left ( t \E [Y] \right ) & \leq \sum_\ell exp \left (t  \sum_{j=1}^k \mu_{\ell,j} + \frac{1}{2} \left ( \sum_{j=1}^k \sigma_{\ell,j}^2 \right ) t^2 \right ), \\
%& = L \sum_\ell \frac{1}{L} exp \left (t  \sum_{j=1}^k \mu_{\ell,j} + \frac{1}{2} \left ( \sum_{j=1}^k \sigma_{\ell,j}^2 \right ) t^2 \right ), \\
& \leq L exp \left (  t   \mu + \frac{\sigma^2}{2} t^2 \right ),
\end{align*}

\noindent where $\mu= \max_\ell \sum_{j=1}^k \mu_{\ell,j}$ and $\sigma^2=\max_\ell \sum_{j=1}^k \sigma_{\ell,j}^2$.  By taking the logarithm on both sides, we thus have $\E [Y] \leq \frac{1}{t} \log(L) + \mu +  \frac{\sigma^2}{2} t$ for all $t>0$. As the minimum of the right-hand side is obtained for $t=\sqrt{2 \log(L)}/\sigma$, we obtain $\E [Y] \leq \mu + \sqrt{2 \log(L)}\sigma$, that is,

\begin{equation}
\E \left [\max_\ell \sum_{j=1}^k P_{\ell,j} \right ] %& \leq \frac{1}{L} \sum_\ell \sum_{j=1}^k \mu_{\ell,j} + \sqrt{\frac{2 \log(L)}{L}} \sqrt{\sum_\ell \sum_{j=1}^k \eta_{\ell,j}^2 \mu_{\ell,j}^2}. \label{eq:bound-on-bottleneck-a} \\
 \leq \max_\ell \sum_{j=1}^k \mu_{\ell,j} + \sqrt{2 \log(L)} \max_\ell  \sqrt{\sum_{j=1}^k \eta_{\ell,j}^2 \mu_{\ell,j}^2}. \label{eq:bound-on-bottleneck-b} 
\end{equation}

It yields

\begin{align}
\E \left [\max_\ell \sum_{j=1}^k P_{\ell,j} \right ] & \leq  \max_\ell \sum_{j=1}^k \mu_{\ell,j} + \sqrt{2 \log(L)} \, \eta_{max} \, \max_\ell  \sqrt{\sum_{j=1}^k \mu_{\ell,j}^2}, \label{eq:bound-on-bottleneck-c} \\
& \leq  \max_\ell \sum_{j=1}^k \mu_{\ell,j} + \sqrt{2 \log(L)} \, \eta_{max} \, \max_\ell \sum_{j=1}^k \mu_{\ell,j}, \label{eq:bound-on-bottleneck-d}
\end{align}

\noindent so that $\alpha \leq 1+ \sqrt{2 \log(L)} \, \eta_{max}$, as claimed.
\end{proof}

\end{document}